\documentclass[preprint]{aastex63}

\usepackage{xcolor}
\usepackage{enumerate}

\let\a=\alpha \let\b=\beta \let\g=\gamma \let\d=\delta \let\e=\epsilon
\let\h=\eta  \let\k=\kappa \let\l=\lambda 
\let\s=\sigma \let\t=\tau \let\u=\upsilon
\let\r=\rho
 \let\eps=\varepsilon
\let\w=\omega	   \let\G=\Gamma \let\D=\Delta  \let\L=\Lambda
\let\S=\Sigma \let\W=\Omega
\let \x = \chi \let \z = \xi

\def\size{\displaystyle}

\def\be{\begin{eqnarray}}
\def\ee{\end{eqnarray}}
\def\ba{\begin{array}}
\def\ea{\end{array}}

\def \Msun {M_{\odot}}
\def \erg {{\rm erg}}

\def \kms {\mathrm{km \, s^{-1}}}
\def \cm {\mathrm{cm}}
\def \M  {\mathscr{M}}
\def \K {\mathrm{K}}
\def \sec {\mathrm{s}}
\def \eV {\mathrm{eV}}
\def \Hz {\mathrm{Hz}}
\def \EM {\mathrm{EM}}

\def \S {Section }

\def \size {\displaystyle}
\def \no {\nonumber}

\shorttitle{EUV spectrum from mm recombination lines}
\shortauthors{Murchikova et al}

\graphicspath{{./}{figures/}}

\begin{document}

\title{Reconstructing EUV spectrum of star forming regions from millimeter recombination lines of  HI, HeI, and HeII}

\correspondingauthor{Lena (Elena M.) Murchikova}
\email{lena@ias.edu}

\author[0000-0001-8986-5403]{Lena Murchikova},
\affiliation{Institute for Advanced Study, 1 Einstein Drive, Princeton, NJ 08540, USA}

\author[0000-0001-7089-7325]{Eric J. Murphy}
\affiliation{National Radio Astronomy Observatory 520 Edgemont Road Charlottesville, VA 22903}

\author[0000-0002-0500-4700]{Dariusz C. Lis}
\affiliation{Jet Propulsion Laboratory, California Institute of Technology, 4800 Oak Grove Drive, 
Pasadena, CA 91109, USA}

\author[0000-0003-3498-2973]{Lee Armus}
\affiliation{Infrared Processing and Analysis Center, MC 314-6, 1200 E. California Boulevard, Pasadena, CA 91125, USA}

\author[0000-0001-9336-2825]{Selma de Mink}
\affiliation{Anton Pannekoek Institute for Astronomy, University of Amsterdam, Science Park 904, 1098XH, \\Amsterdam, The Netherlands}
\affiliation{Center for Astrophysics, Harvard Smithsonian, 60 Garden Street, Cambridge, MA 02138, USA}

\author[0000-0002-5496-4118]{Kartik Sheth}
\affiliation{NASA Headquarters, 300 E Street SW, Washington, DC 20546, USA}

\author[0000-0001-6100-6869]{Nadia Zakamska}
\affiliation{Department of Physics and Astronomy, Johns Hopkins University, 3400 N. Charles St., Baltimore, MD 21218, USA}

\author[0000-0001-8631-7700]{Frank Tramper}
\affiliation{Institute for Astronomy, Astrophysics, Space Applications and Remote Sensing, National Observatory of Athens,\\ Metaxa \& Vas. Pavlou St., 15236, Penteli, Athens, Greece}

\author{Angela Bongiorno}
\affiliation{INAF-Osservatorio Astronomico di Roma, Via di Frascati 33, I-00040 Monteporzio Catone, Rome, Italy}

\author{Martin Elvis}
\affiliation{Center for Astrophysics, Harvard Smithsonian, 60 Garden Street, Cambridge, MA 02138, USA}

\author{Lisa Kewley}
\affiliation{RSAA, Australian National University, Cotter Road, Weston Creek, ACT 2611, Australia}

\author{Hugues Sana}
\affiliation{Institute of Astrophysics, Universiteit Leuven, Celestijnenlaan 200 D, 3001 Leuven, Belgium}

%\nocollaboration{2}

\begin{abstract}

The extreme ultraviolet (EUV) spectra of distant star-forming regions cannot be probed directly using either ground- or space-based telescopes due to the high cross-section for interaction of EUV photons with the interstellar medium. This makes EUV spectra poorly constrained. The mm/submm recombination lines of H and He, which can be observed from the ground, can serve as a reliable probe of the EUV. Here we present a study based on ALMA observations of three Galactic ultra-compact HII regions and the starburst region Sgr B2(M), in which we reconstruct the key parameters of the EUV spectra using mm recombination lines of HI, HeI and HeII. We find that in all cases the EUV spectra between 13.6 and 54.4 eV have similar frequency dependence: $L_{\nu}\propto \nu^{-4.5 \pm 0.4}.$ We compare the inferred values of the EUV spectral slopes with the values expected for a purely single stellar evolution model (Starburst99) and the Binary Population and Spectral Synthesis code (BPASS). We find that the observed spectral slope differs from the model predictions. This may imply that the fraction of interacting binaries in HII regions is substantially lower than assumed in BPASS. The technique demonstrated here allows one to deduce the EUV spectra of star forming regions providing critical insight into photon production rates at $\l \leq 912 \mathring{\mathrm{A}}$ and can serve as calibration to starburst synthesis models, improving our understanding of star formation in distant universe and the properties of ionizing flux during reionization.

\end{abstract}


\section{Introduction} \label{sec:intro}

Wherever ionization is taking place extreme ultraviolet radiation (EUV) $\sim 10 \eV - 100 \eV$ is being emitted and absorbed by gas and dust. However, EUV radiation is notoriously difficult to observe. 
Due to their large cross section for interaction with the interstellar medium, EUV photons 
can typically travel no more than several hundred parsecs \citep{2006ASPC..352...79R} making direct observations of EUV spectra of distant star forming galaxies impossible. Even for nearby sources, the EUV observations are difficult because EUV photons are easily absorbed by the Earth's atmosphere, they can only be observed from space.
Instruments in space are currently quite limited in their wavelength coverage and spectral resolution \citep{2010LanB...4A..109W}, and therefore accurate indirect methods of determining the EUV fluxes of astronomical sources are very valuable. 

Young stellar populations containing massive stars largely determine the radiative properties of star-forming galaxies and are the primary source of ionizing radiation, in the absence of an active central supermassive black hole, e.g. \citet{1981PASP...93....5B, 1999ApJS..125..489G, 2012ApJ...752..162S}. Optical emission lines, e.g. through BPT diagrams \citep{1981PASP...93....5B,1987ApJS...63..295V}, are often used to indicate the presence of an underlying star forming population. The total production rates of ionizing EUV photons by young star clusters are often determined from visible wavelength H$\a$ and H$\b$ emission line fluxes or from radio free-free continuum emission (e.g. \citet{2012ApJ...761...97M,2020arXiv200410230L,2017ApJ...839...35M}). 
As a result, quantitative relationships between line fluxes and the underlying EUV fluxes are often made difficult by extinction uncertainties. 

The uncertain EUV fluxes are of large concern for the widely used spectral synthesis codes such as Starburst99 \citep{1999ApJS..123....3L,2014ApJS..212...14L}. Recently, stellar evolutionary models have been improved by more realistic treatment of mass loss, stellar rotation, binarity and the upper stellar mass limit \citep{2000PhDT........45V, 2000A&A...361..159M, 2009A&A...507L...1D, 2016MNRAS.458..624C}. This new generation of models, however, shows large discrepancies in the extreme UV \citep{Levesque12,Stanway16,2017PASA...34...58E} 
due to stellar wind mass-loss, model atmosphere uncertainties (see Fig. 4 of \citet{2012ASInC...6...79L}) and most importantly, whether there are Wolf-Rayet stars present. The latter, being very hot ($\sim 50,000-200,000$ K) and of high luminosity, can increase the hard EUV luminosity by 2-4 orders of magnitude. The incidence of Wolf-Rayet stars is affected by assumed stellar rotation and binarity and by the adopted upper mass limit.

Empirical methods are needed to reconstruct accurate EUV properties of star formation to bring us closer to understanding finer detail of star formation, and to provide critical input for testing and constraining population synthesis models. This in turn impacts cosmological and galaxy formation simulations, which rely on synthesis codes for treatment of subgrid physics input: ionizing fluxes and radiative feedback, as well as for the interpretations of the observations of galaxies in the local Universe and at high redshift.

Such a probe of EUV spectra can be provided by the millimeter/submm recombination lines of HI, HeI and HeII. These lines can probe the EUV spectrum at energies $>13.6 \, \eV,$ $>24.6 \, \eV$ and $>54.4 \, \eV,$ corresponding to the ionization threshold of HI, HeI and HeII, respectively. These lines circumvent all of the problems with other indirect probes: (1) they are at sufficiently long wavelengths that the dust opacity should be negligible; (2) they are permitted transitions with high critical densities (populated by recombination and radiative decay); (3) they do not have significant maser amplification (an issue for cm-wave recombination lines); and (4) they are emitted by H and He and thus do not depend on metallicity \citep{2013ApJ...779...75S} [hereafter SM13]. 

The theoretical analysis of SM13 indicates that the mm/submm line fluxes provide reliable estimates of the emission measures of HII and HeIII, which we extend here to HeII. Using these emission measures one can then estimate the ionizing continua of HI, HeI and HeII, assuming radiative equilibrium. Since these lines are ionized at different photon energies, by comparing the ionizing continua of these three species one can derive the underlying EUV spectrum. 
We are particularly interested in the so called $\alpha$-transitions, $\rm{n}+1 \to \rm{n}$ transitions of neutral hydrogen H, neutral He and of single-ionized He (see SM13). Over the density range $n_e = 10^2-10^8 \, \cm^{-3},$ the emissivities of the $\a$ lines of HI, HeI and HeII at $\mathrm{n}\sim 30$ vary by less than 20\%, and variations with $T_e$ are also small. %SM13 also determined that the mm/submm lines of both species have positive opacity coefficients and hence do not have maser amplification.

In this work we report the result of an Atacama Large Millimeter/submillimeter Array (ALMA) observing program %pilot program 
demonstrating the potential of mm/submm lines of HI, HeI and HeII to reveal the underlying EUV spectrum. To this end we use a sample of three ultra-compact HII regions and one massive starburst region (i.e. a region containing multiple HII regions). The intrinsic dusty environment of compact star forming regions makes observations in mm/submm lines particularly advantageous.

The paper is organized as follows: in Section \ref{sec:theory} we discuss the theory of using the recombination lines of HI, HeI and HeII to constrain the EUV spectra. In Section \ref{sec:data1} we describe the observations and data reduction. In Section \ref{sec:data2} we discuss the data analysis. In Section \ref{sec:results} we discuss the physical implication of the obtained EUV spectra and compare them with simulations. We use BPASS v2.2.1 and Starburst 99 v7.0.1. In Appendix \ref{app:ion} we calculate the production rate of EUV photons and derive the properties of EUV spectra.

\bigskip

\section{Theory}
\label{sec:theory}

In this section we derive the intrinsic EUV spectra inside a star forming (SF) region using observed mm recombination lines fluxes of H30$\a$, He30$\a$ and He$^+$48$\a.$ Any three mm/submm recombination lines of HI, HeI and HeII may be used, but these three particular lines are chosen here because they can be observed simultaneously with ALMA, and thus are favorable from the point of view of reducing the telescope time. Below we first convert the observed velocity-integrated line fluxes into a volume emission measure, then, assuming radiation equilibrium, derive the production rates of EUV photons, and then derive the EUV spectral parameters. 

The velocity-integrated line flux $[S\D V]_\mathcal{L}$ in mm/submm recombination line $\mathcal{L}$ from a SF region is directly proportional to the volume emission measure $\EM_{x}$ of the ion $x$ which produces the line $\mathcal{L}$ while recombining with an $e^-$ through recombination cascade (SM13):
\be\label{sdv}
	[S\D V]_\mathcal{L} = \frac{\eps_\mathcal{L}}{4 \pi D^2} \, \mathrm{EM}_x \frac{c}{\nu_{\mathcal{L}}}.
\ee
Here $[S\D V]_\mathcal{L} =\int S_\mathcal{L}(V) dV $ is the velocity-integrated line flux, $S_\mathcal{L}(V)$ is the flux in the $\mathcal{L}$ line at velocity $V$, $\eps_\mathcal{L}$ is the emissivity of the recombination line $\mathcal{L},$ $D$ is the distance to the source,
$c$ is the speed of light, $\nu_{\mathcal{L}}$ is the frequency of line $\mathcal{L}.$ The volume emission measure is defined as
$\mathrm{EM}_{x}= n_{x} n_{e} \, vol_x,$ $n_x$ is density of ion $x,$ $n_e$ is electron density in the region,
and $vol_x$ is volume of space taken by the region where species $x$ is present. In our observations we specifically consider $x=$HII, HeII, HeIII and $\mathcal{L} = \{\mathrm{H30\a: \, n}=31 \to 30\},$ $\{\mathrm{He30\a: \, n}=31 \to 30\},$ $\{\mathrm{He^+48\a: \, n}=49 \to 48\}$.

The line frequencies and emissivities of the observed recombination lines H30$\a,$ He30$\a,$
and He$^+48\a$ at typical densities $n\sim 10^4 \, \cm^{-3}$ and temperatures $T\sim10^{4}$ K inside HII regions are \citep{sto95a,sto95}: 
\be
\no\size {\rm H}30\a: & \quad    \nu_\mathrm{H30\a}=231.901 \, {\rm GHz} , \quad & \eps_\mathrm{H30\a}=1.05 \times 10^{-31}{\erg \, \cm^3 \, \sec^{-1}} \\
\label{eq:lines} \size {\rm He}30\a: & \quad   \nu_\mathrm{He30\a}=231.995 \, {\rm GHz}, \quad &  \eps_\mathrm{He30\a}=1.05 \times 10^{-31}{\erg \, \cm^3 \, \sec^{-1}} \\
\no\size {\rm He}^+48\a: & \quad \nu_\mathrm{He^+48\a}=230.713 \, {\rm GHz} ,  \quad & \eps_\mathrm{He^+48\a}=3.67 \times 10^{-31}{\erg \, \cm^3 \, \sec^{-1}}. 
\ee
While the values of $\eps_\mathrm{H30\a}$ and $\eps_\mathrm{He^+48\a}$ are tabulated in the literature, to estimate the value of $\eps_\mathrm{He30\a}$ we use the fact that at $\mathrm{n} \sim 30$ HeI atom can also be considered a hydrogenic ion as the second electron is close to the nucleus and is screening its charge. Then we can use scaling relation $\eps_{\mathrm{He}\, \mathrm{n}\a}(T,n) \simeq Z_\mathrm{eff}^{17/3} \eps_{\mathrm{H}\mathrm{n}\a}(T,n)$ derived in SM13 Appendix C and assume $Z_\mathrm{eff}= 1.0002$ for the effective charge of the nucleus \citep{Towle96}.

From eq. (\ref{sdv}) and (\ref{eq:lines}) we find
\be\label{SdVtoEM}
	\left[\ba{l}
	\frac{EM_{\rm HII}}{\cm^{-3}}\\[4pt]
	\frac{EM_{\rm HeII}}{\cm^{-3}}\\[4pt]
	\frac{EM_{\rm HeIII}}{\cm^{-3}}
	\ea\right]
	=\left[\ba{c} 
	8.81\\[4pt]
	8.81\\[4pt]
	2.51
	\ea
	\right]\times 10^{57}
	\left[\ba{l} 
	\frac{{[S\D V]}_{\rm H30\a}}{\mathrm{Jy} \, \kms}\\[4pt]
	\frac{{[S\D V]}_{\rm He30\a}}{\mathrm{Jy} \, \kms}\\[4pt]
	\frac{{[S\D V]}_{\rm He^{+}48\a}}{\mathrm{Jy} \, \kms}
	\ea
	\right]	\left( \frac{D}{{\rm kpc}} \right)^2.
\ee

To keep the amount of gas constituting the EMs ionized we need an influx of ionizing photons above the ionization threshold ($13.6 \eV$, $24.6 \eV$ and $54.4 \eV$ for HI, HeI and HeII respectively) sufficient to counteract recombination. Then, following Appendix \ref{app:ion}, we can use the following approximate relations
\be \label{eq:EMtoQ}
	\left[\ba{l}
	\frac{Q_0}{\sec^{-1}}\\[3pt]
	\frac{Q_1}{\sec^{-1}}\\[3pt]
	\frac{Q_2}{\sec^{-1}}
	\ea\right]
	\simeq\left[\ba{c} 
	2.59\\[3pt]
	2.72\\[3pt]
	18.5
	\ea
	\right]\times 10^{-13}
	\left[\ba{l} 
	\frac{{\rm EM}_\mathrm{HII}}{\cm^{-3}}\\[3pt]
	\frac{{\rm EM}_\mathrm{HeII}}{\cm^{-3}}\\[3pt]
	\frac{{\rm EM}_\mathrm{HeIII}}{\cm^{-3}}
	\ea
	\right].
\ee  
Here we use the standard notations for the photon production rates above the HI, HeI and HeII ionization thresholds
\be\label{Qs}
 Q_0= \int^{\infty}_{\nu_0} \frac{L_\nu}{h \nu} {d \nu}, \qquad 
 Q_1= \int^{\infty}_{\nu_1} \frac{L_\nu}{h \nu} {d \nu}, \qquad 
 Q_2= \int^{\infty}_{\nu_2} \frac{L_\nu}{h \nu} {d \nu},
\ee
where $L_\nu$ is the specific luminosity of all the stars within the region and $h\nu_0=13.6\eV,$ $h\nu_1=24.6\eV$ and $h\nu_2 = 54.4\eV.$

Now we need to assume a model for the EUV spectrum inside SF regions. Due to the effects of stellar atmospheres it is natural to approximate the EUV spectrum for photons with energies higher than 13.6 eV as a three-component places power-law. The first break at 24.6 eV corresponds to the ionization threshold of HeI and the second break at 54.4 eV to the ionization threshold of HeII. Such a model however has four free parameters -- the absolute value of the spectra at 13.6 eV, the spectral slope between 13.6 eV and 24.6 eV, the spectral slope between 24.6 eV and 54.4 eV, and the spectral slope above 54.4 eV. Our observations provide us with three observed parameters, see equations \ref{SdVtoEM}, and thus an ability to determine three model parameters. Consequently we need to eliminate one of the free model parameters. We further assume that the EUV spectrum inside SF regions is well-described by a characteristic broken power-law with two distinct spectral slopes, the first of which is $\gamma,$ the EUV spectral slope between energies 13.6 eV and 54.4 eV, and the second is $\gamma_2$ -- the slope for energies higher than 54.4 eV. Thus we find (see Appendix \ref{app:ion})
\be\label{eq:lnu}
 L_\nu = \left\{
 \ba{ll}
  \size 1.716 \times 10^{22} \,  {\erg \, }{\sec^{-1} \, \Hz^{-1}} \left(\frac{{\rm EM}_{\rm HII}}{10^{61} \, \cm^{-3}}\right) \frac{\gamma}{1-0.25^{\gamma}} \left(\frac{\nu}{\nu_0}\right)^{-\gamma},    & \quad \nu_0 \leq \nu < \nu_2 \\
 \size L_2 (\nu/\nu_2)^{-\gamma_2},& \quad \nu \geq \nu_2,
  \ea
 \right. 
\ee 
where 
\be
\gamma_2 \simeq 1.4 \times 10^{2} \left(\frac{{\rm EM}_{\rm HII}}{10^{61} \, \cm^{-3}}\right) \left(\frac{{\rm EM}_{\rm HeIII}}{10^{58} \, \cm^{-3}}\right)^{-1} \frac{\gamma}{4^\gamma-1},
\ee 
$\gamma$ is the solution of the equation
\be  
    \frac{2.21^{\gamma}-1}{4^{\gamma}-1} \simeq 0.105 \left(\frac{{\rm EM}_{\rm HeII}}{10^{60} \, \cm^{-3}}\right) \left(\frac{{\rm EM}_{\rm HII}} {10^{61} \, \cm^{-3}}\right)^{-1}.
\ee  
and from the continuity requirement we get
\be
 L_2 = 1.716 \times 10^{22} \frac{\erg}{\sec \, \Hz}  \left(\frac{{\rm EM}_{\rm HII}}{10^{61} \, \cm^{-3}}\right) \frac{\gamma}{4^\gamma-1}.
\ee

\bigskip

\section{Observations and Data reduction}
\label{sec:data1}

%%%%%
We present results from ALMA observations. 
Our data were obtained during two ALMA observational programs: Cycle 2 project 2013.1.00111.S and Cycle 4 project 2016.1.01015.S. Table \ref{tab:targets} lists the names of the observed HII regions, their coordinates, the distances to the sources and ALMA Cycle number in which the data were taken. 

For the initial set of observations in ALMA Cycle 2, the target list included four OB star formation regions in our Galaxy: three ultra-compact HII regions G330.9536, G332.8254 and G328.8076 and the very massive starburst Sgr B2(M), which consists of several HII regions. The three ultra-compact HII regions are taken from the survey of \cite{2010MNRAS.405.1560M}. They were selected for their high luminosity and compactness as well as their location in the southern hemisphere to be most convenient for ALMA observations. Compactness is desired observationally, so 
there would be no concerns about needing extra baselines, and thus observations in additional configurations, to recover extended faint structure,, and physically so that the ionized regions are likely to be ionization bounded with most of the EUV continuum absorbed.

%%%%%

\begin{deluxetable*}{lrrrcc}
\tabletypesize{}
\tablecaption{Observed HII regions} \label{tab:targets}  
\tablewidth{0pt}
\tablehead{
\colhead{Name } & \colhead{RA } & \colhead{DEC} & \colhead{Distance} & \colhead{Ang. scale} & \colhead{ALMA Cycle}
}
\startdata
G330.9536 & 16:09:52.60 & -51:54:55 & 5.5 kpc & $\sim$ 2'' & 2 \& 4 \\
G332.8254 & 16:20:11.00 & -50:53:16 & 4.4 kpc & $\sim$ 3'' & 2 \& 4 \\
SgrB2(M) & 17:47:20.15 & -28:23:05 & 8.3 kpc & $\sim 2'' \times 4''$ & 2  \\
G328.8076 & 15:55:48.60 & -52:43:07 & 11.7 kpc & $\sim$ 4'' & 2
\enddata
\tablecomments{The list of HII regions selected for this study, their positions, distances, angular scale and the ALMA Cycle they were observed in.}
\end{deluxetable*}

We chose specific recombination lines in this study to avoid confusion with known possible molecular lines by generating model molecular line spectral-scans for the Ori-IRc2 \citep{2014ApJ...787..112C} obtained specifically for this project. 
Among the possible lines, H30$\a,$ He30$\a,$ and He$^+48\a$ lines can be observed simultaneously with ALMA Band 6 making them our preferred choice. Other possible lines include H26$\a$ with the corresponding He26$\a$ and He$^+49\a.$

The three recombination lines H30$\a$ (HI: n=31$\to$30 at 231.901 GHz), He30$\a$ (HeI: n=31$\to$30 at 231.995 GHz) and He$^{+}48\a$ (HeII: n=49$\to$48 at 230.713 GHz) were observed simultaneously in one tuning of ALMA in Band 6 with two spectrometers with a velocity resolution of $2.5 \, \kms.$ The first spectrometer was set to observe the H30$\a$ and the He30$\a$ recombination lines and the second was set to the He$^+$48$\a$ line. The other three spectral windows were set to observe the continuum at 232.8 GHz, 232.3 GHz and 229.9 GHz with a velocity resolution of 40 $\kms.$ The continuum is expected to be a mixture of free-free emission and thermal dust emission. The spectral setup was the same in all of our observations.

The observations in ALMA Cycle 2 were conducted on April 23, 2015 (Sgr B2 (M)) with 38 antennas and achieved a resolution of $1.24'' \times 0.93''$; on April 24 and 27, 2015 (G328.8076, G330.9536, and G332.8254) with 38 antennas and achieved resolution of $0.53'' \times 0.47'',$ and on June 16, 2015 (Sgr B2(M)) with 30 antennas and achieved resolution $0.53'' \times 0.47''.$ The exposure times on target were 9 min on each of the G-sources and 18 minutes total on Sgr B2(M). Achieved sensitivity is 2 mJy/beam in a 10 $\kms$ velocity channel in each source. We used the calibration and data reduction scripts prepared by the staff at North American ALMA Science Center (NAASC) in Charlottesville, Virginia. We imaged and CLEANed the data using the Common Astronomy Software Applications package (CASA). The images created were $500\times 500$ pixel with the pixel size $0.08''.$ In all observed sources the HI recombination line was clearly detected.

For the deeper observations in ALMA Cycle 4 we selected the two most H30$\a$-luminous and 
the sources least contaminated by molecular lines -- G330.9536 and G332.8254.
The observations of G330.9536 and G332.8254 in ALMA Cycle 4 were conducted on May 6 and 8, 2017 with 47 antennas and May 7, 2017 with 50 antennas. The achieved resolution was $\sim 0.3''\times 0.28''.$ The total exposure time on each target was $\sim$120 min.  Achieved sensitivity is 0.4 mJy/beam in a 10 $\kms$ velocity channel in each source. For calibration and data reduction we used the script automatically generated by the ALMA pipeline provided by the NAASC. We imaged the data using the task \verb|tclean| in CASA. The images created were $1000\times 1000$ pixel with the pixel size $0.028''.$ In all the observed sources the HI and HeI recombination lines was clearly detected. We set a limit on HeII recombination line emission. Maximum recoverable scale of the emission is $\sim 3.5''.$

\bigskip

\section{Data analysis}
\label{sec:data2}

\begin{figure*}
\vspace{-0.0cm}
\centering
\begin{tabular}{cc}
\includegraphics[width=0.3\columnwidth]{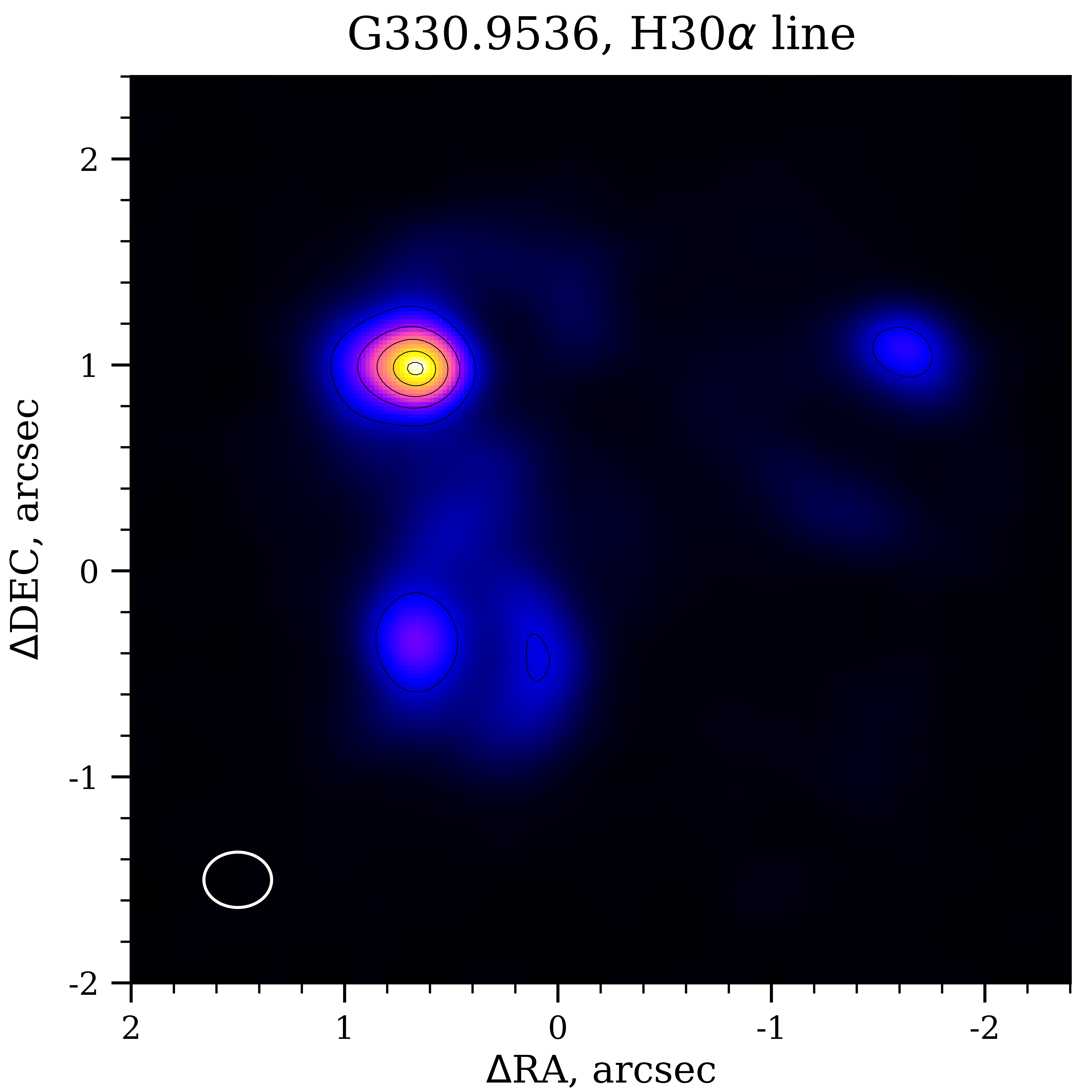} &
\includegraphics[width=0.3\columnwidth]{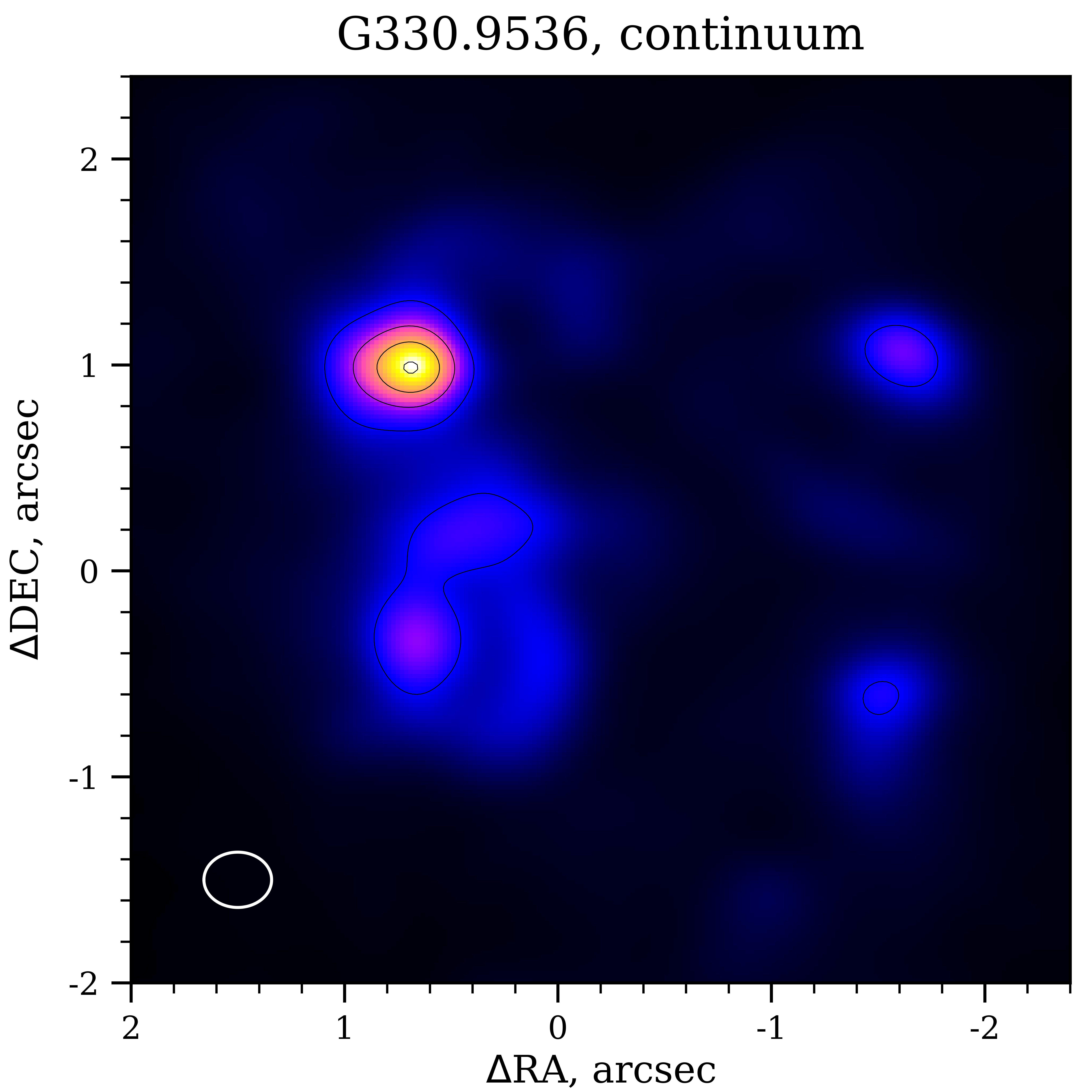} \\
\includegraphics[width=0.3\columnwidth]{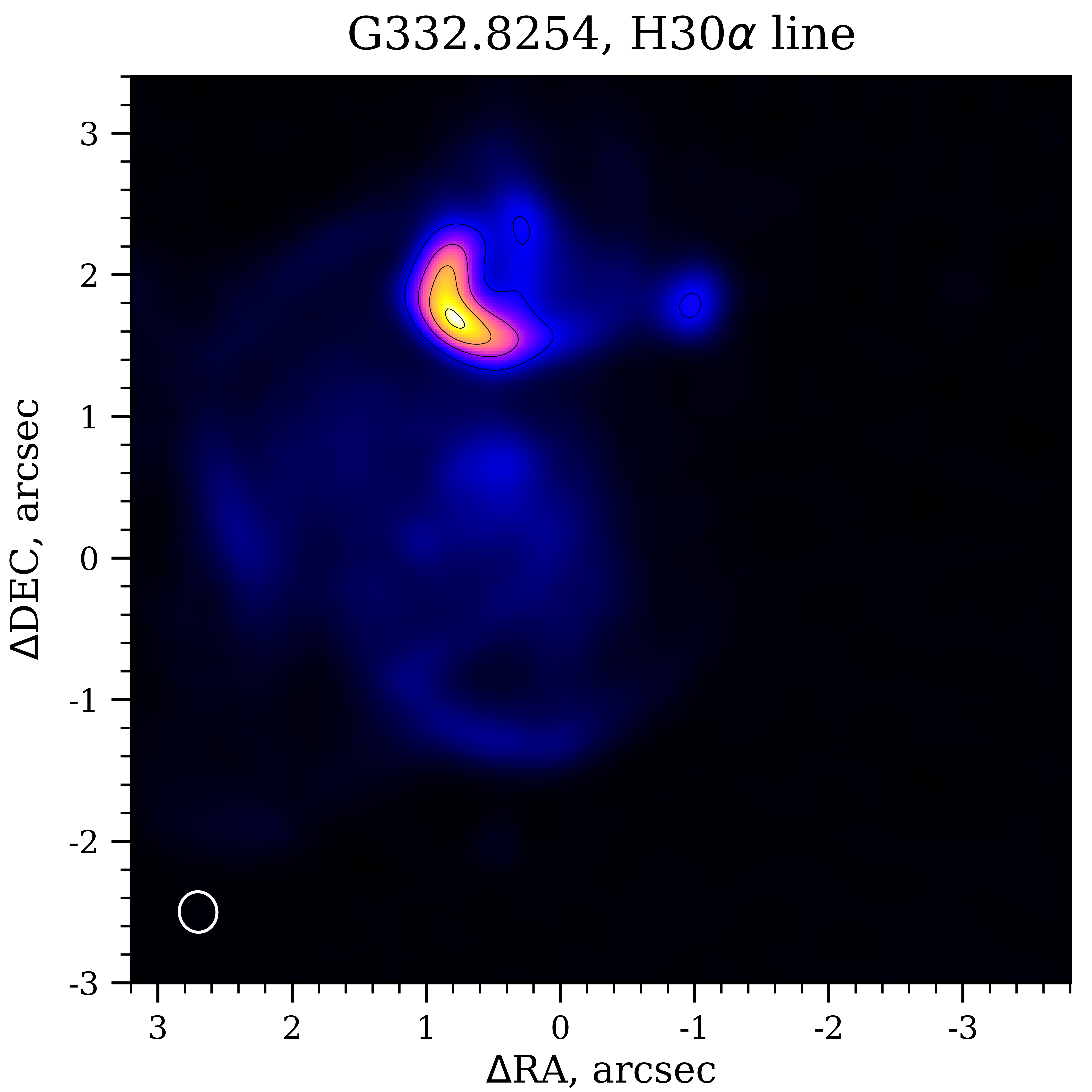} &
\includegraphics[width=0.3\columnwidth]{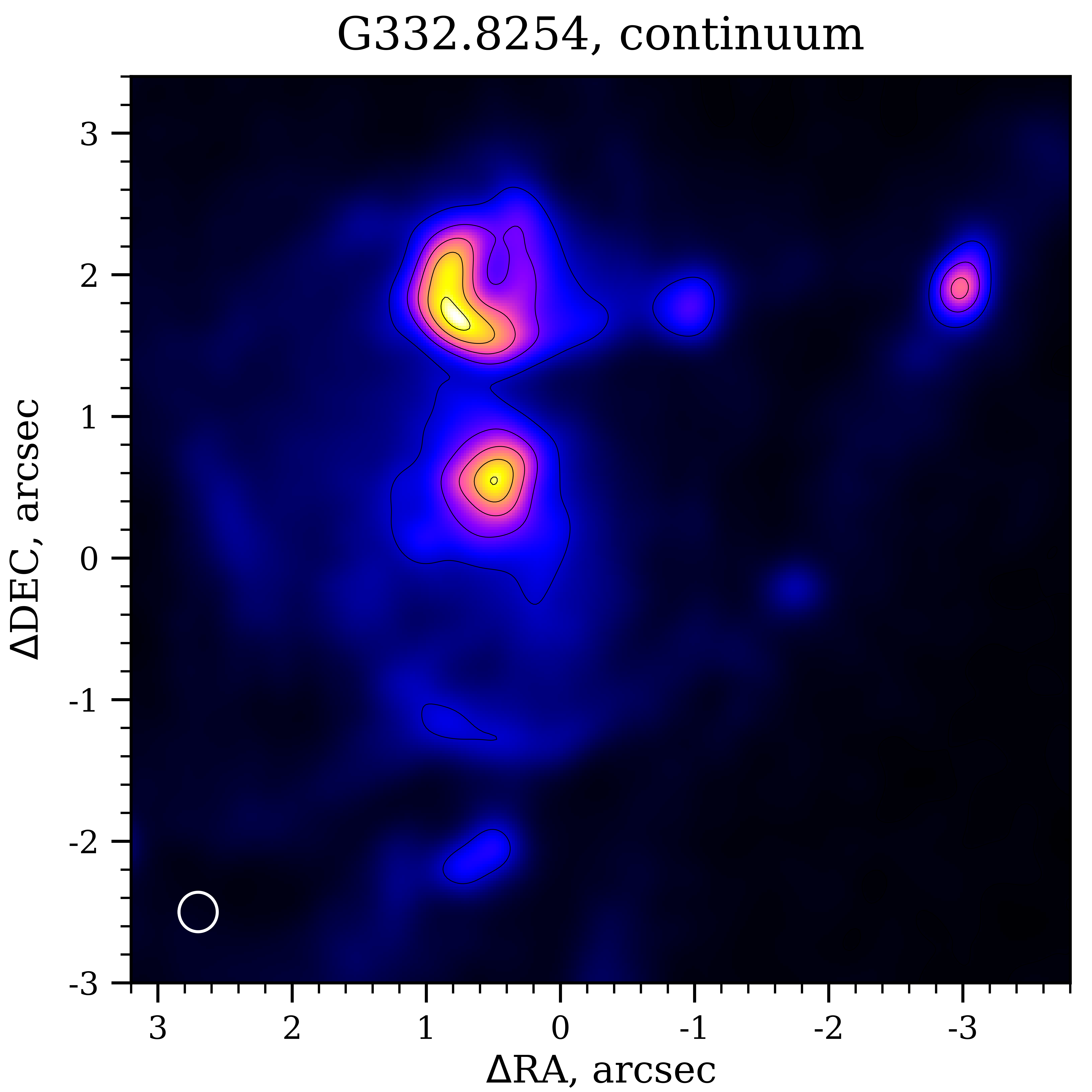} \\
\includegraphics[width=0.3\columnwidth]{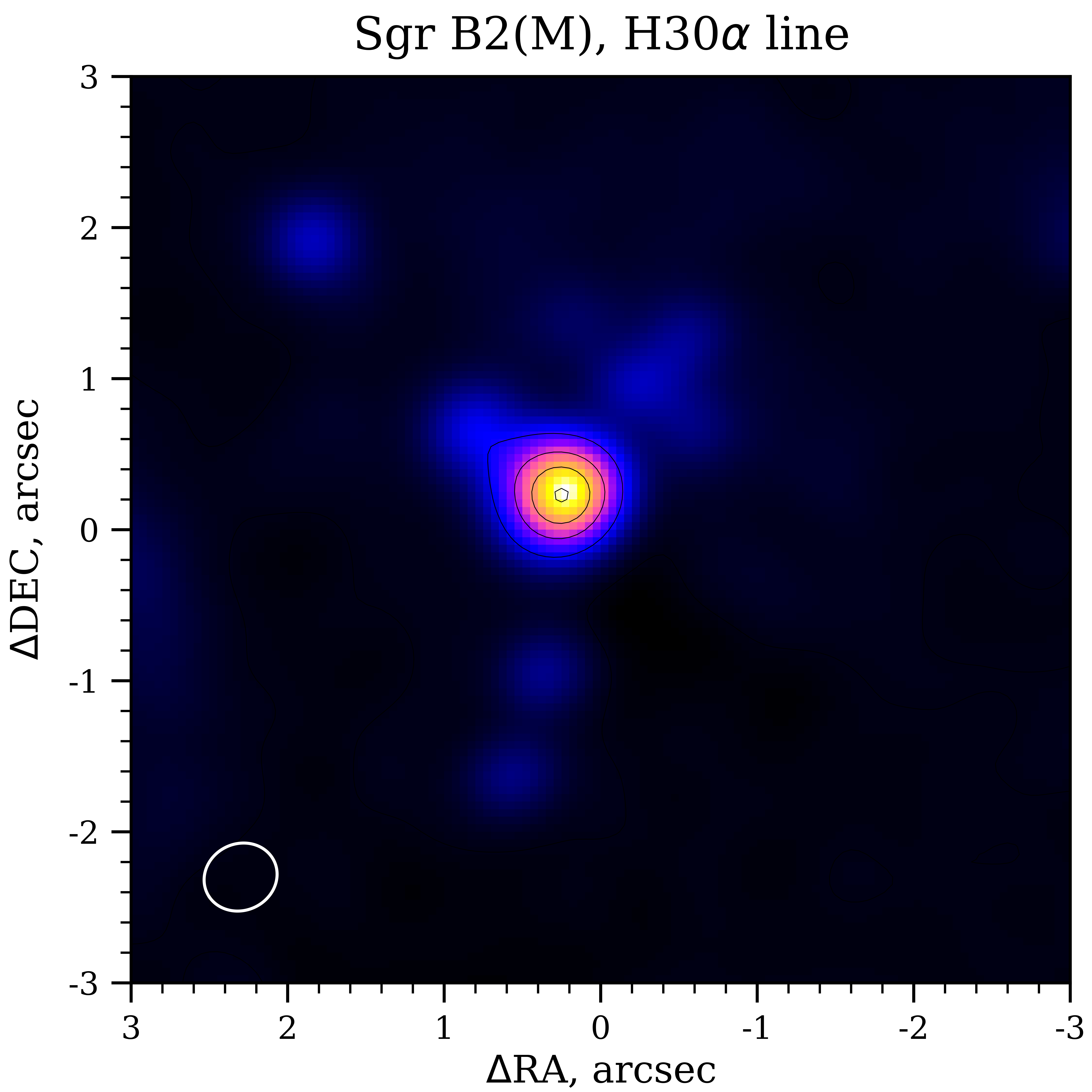} &
\includegraphics[width=0.3\columnwidth]{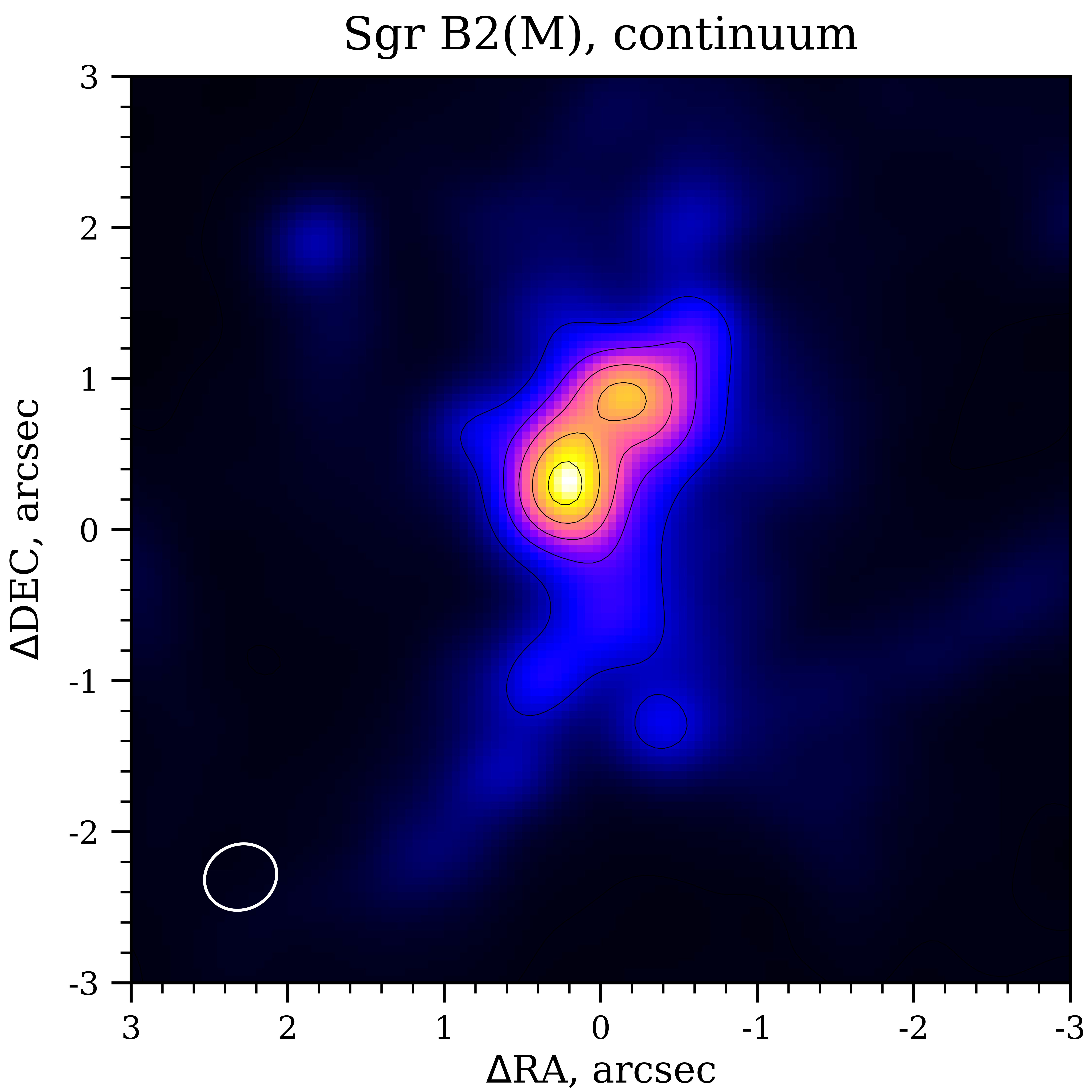} \\
\includegraphics[width=0.3\columnwidth]{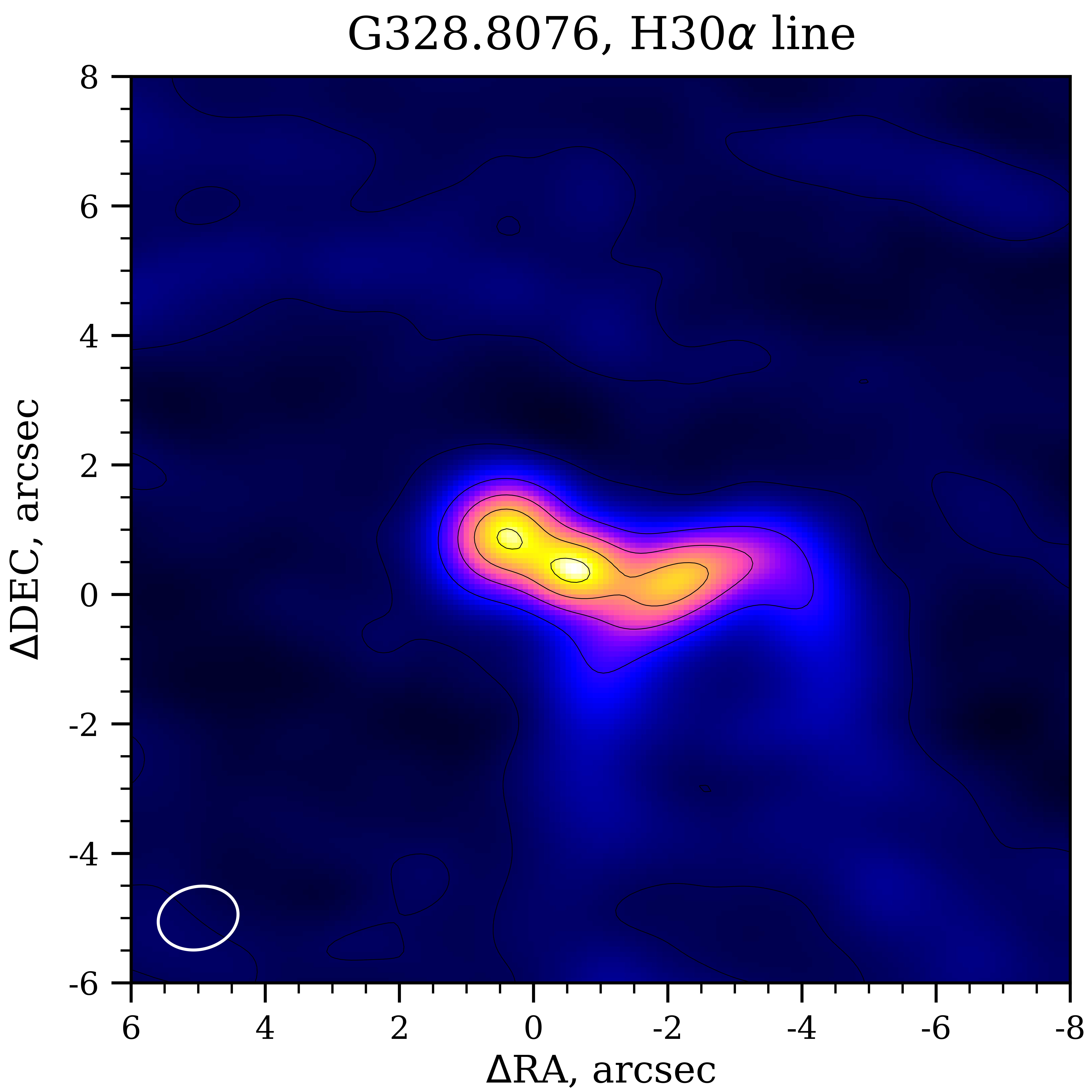} &
\includegraphics[width=0.3\columnwidth]{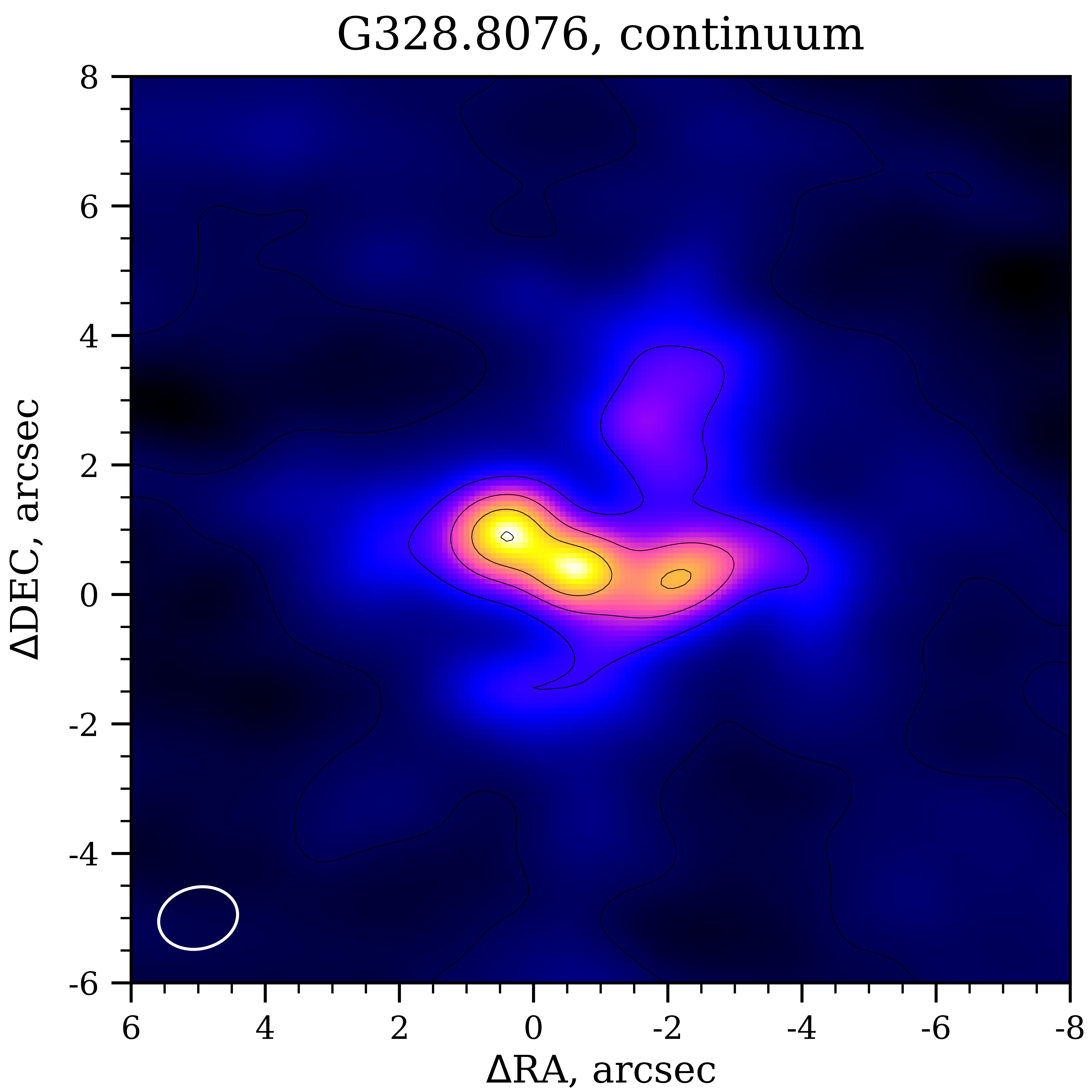} 
\end{tabular}
\caption {Continuum-subtracted H30$\a$ line flux maps (left) and continuum maps (right) of the Galactic HII regions G330.9536, G332.8254, G328.8076 and the massive starburst Sgr B2 (M). The beams sizes are plotted in the bottom left corner of each panel.}
\label{fig:sources}
\end{figure*}

After imaging the data as described above we identified the line-free channels. The HII regions in general have an abundance of molecular and atomic features. Therefore, it is important to identify the channels for continuum subtraction carefully, rather than identifying wide frequency ranges outside the recombination lines of interest. Then we performed subtraction of the continuum in the uv-plane with
the \verb|uvcontsub| task in CASA using the identified channels to obtain separate continuum and line images (Figure \ref{fig:sources}). 

\begin{figure*}
\vspace{-0.0cm}
\centering
\begin{tabular}{cc}
\includegraphics[width=0.48\columnwidth]{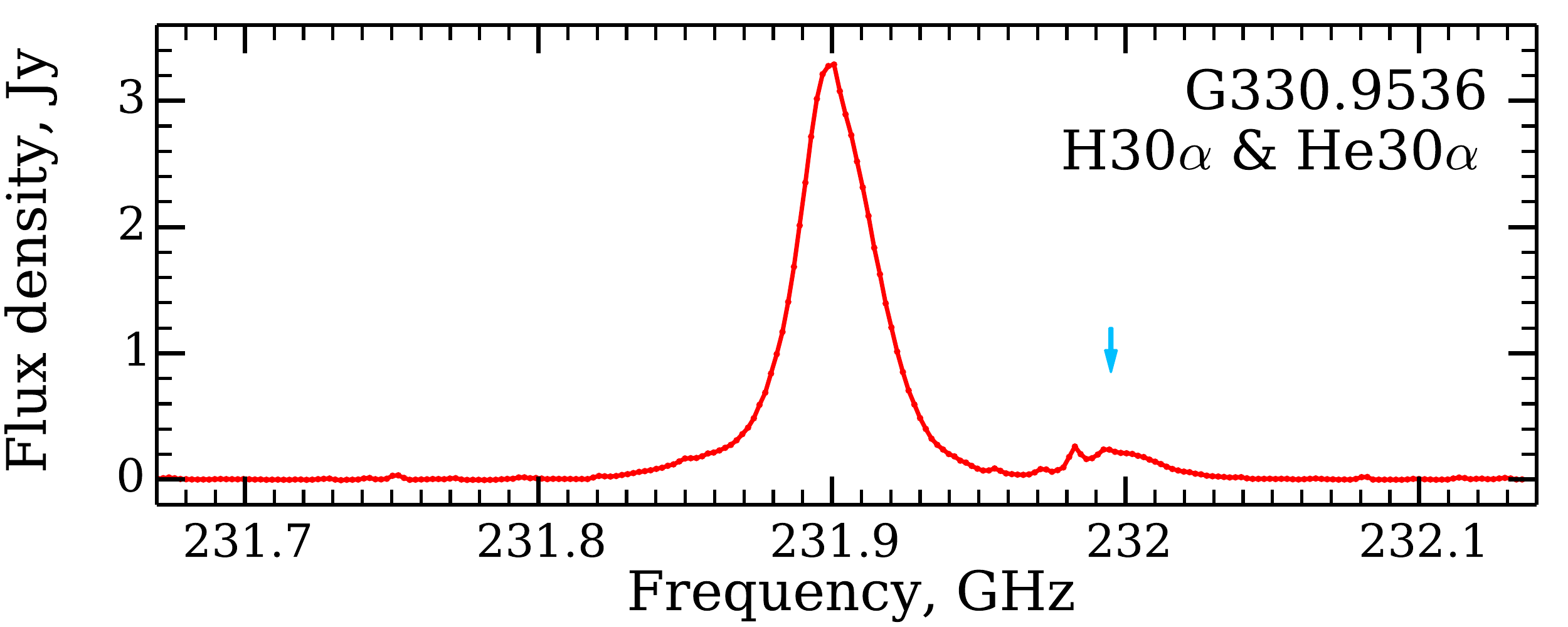} & 
\includegraphics[width=0.48\columnwidth]{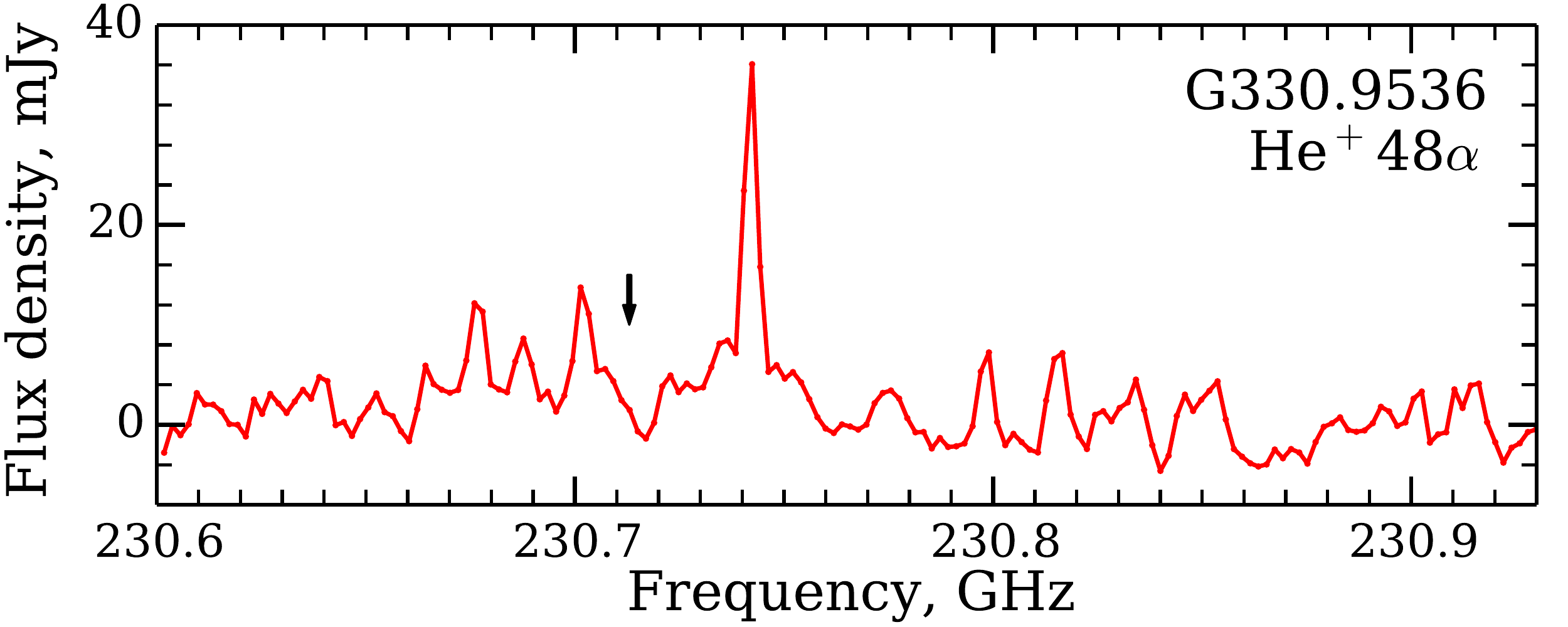}\\
\includegraphics[width=0.48\columnwidth]{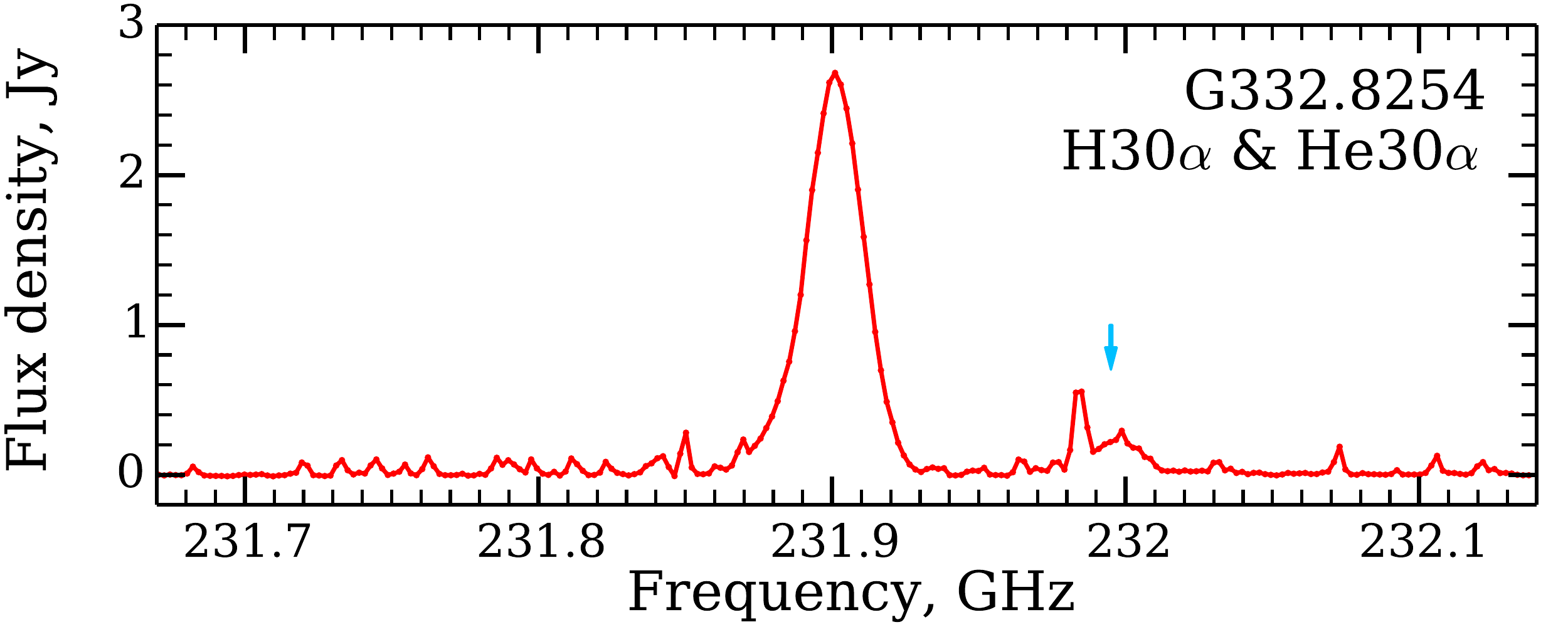} & 
\includegraphics[width=0.48\columnwidth]{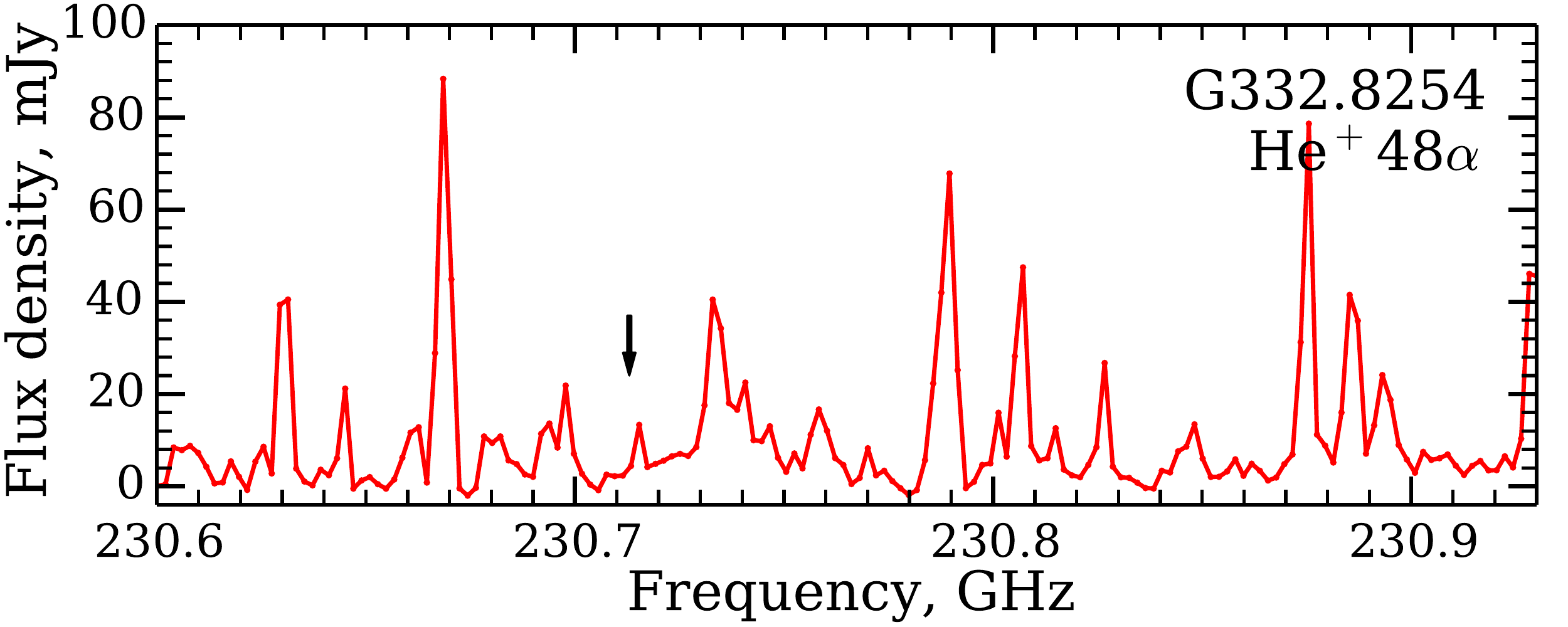}\\
\includegraphics[width=0.48\columnwidth]{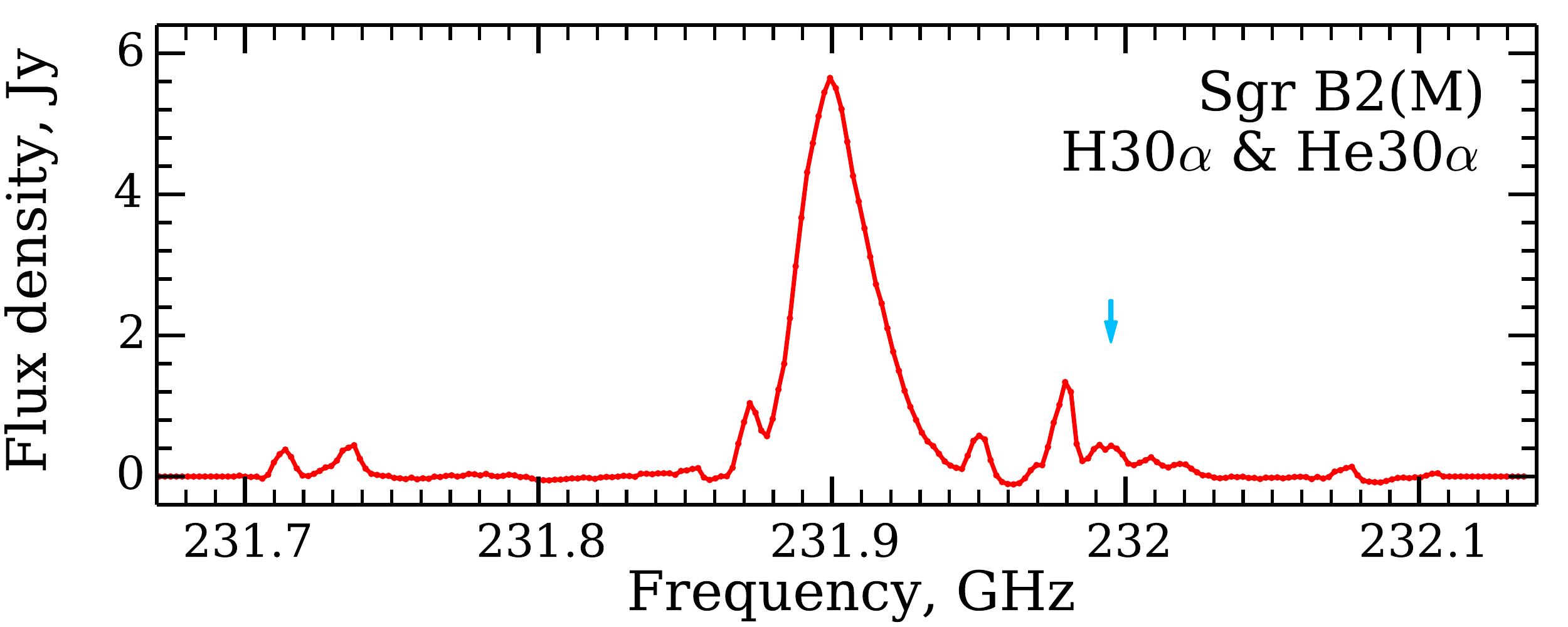} & 
\includegraphics[width=0.48\columnwidth]{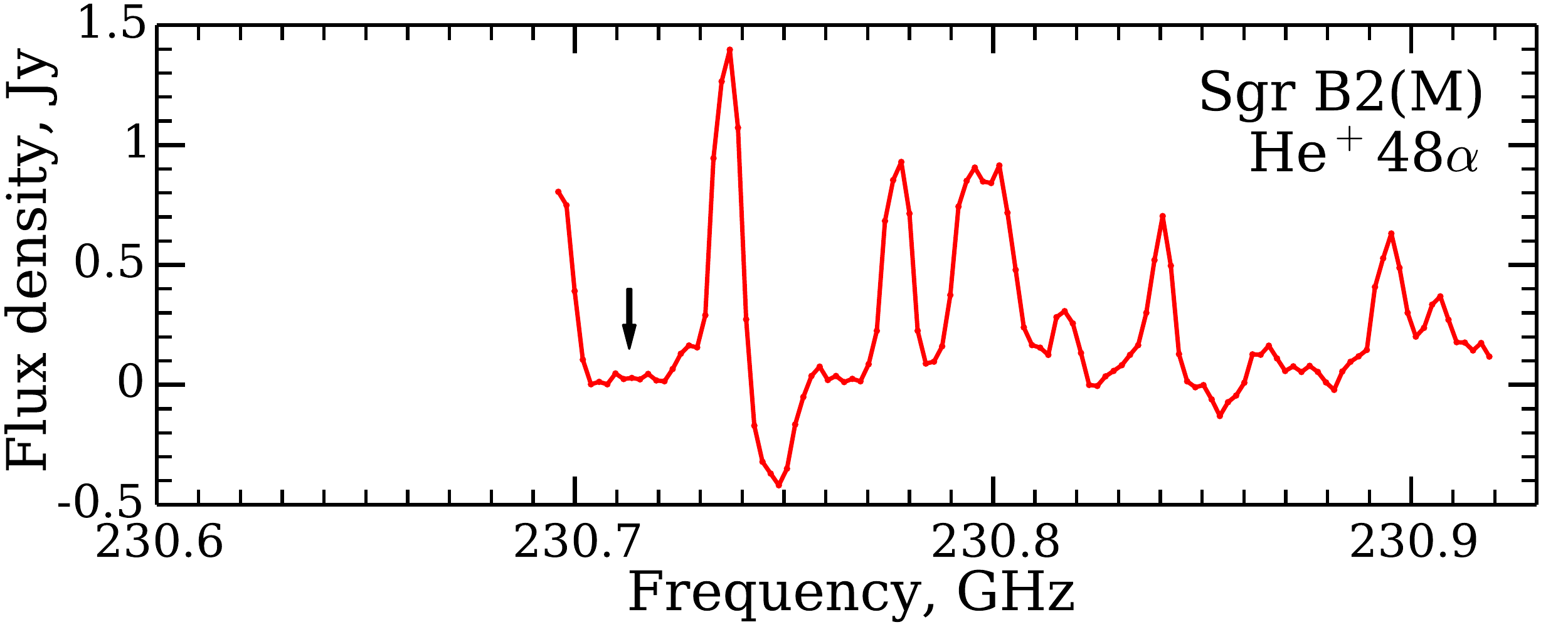}\\
\includegraphics[width=0.48\columnwidth]{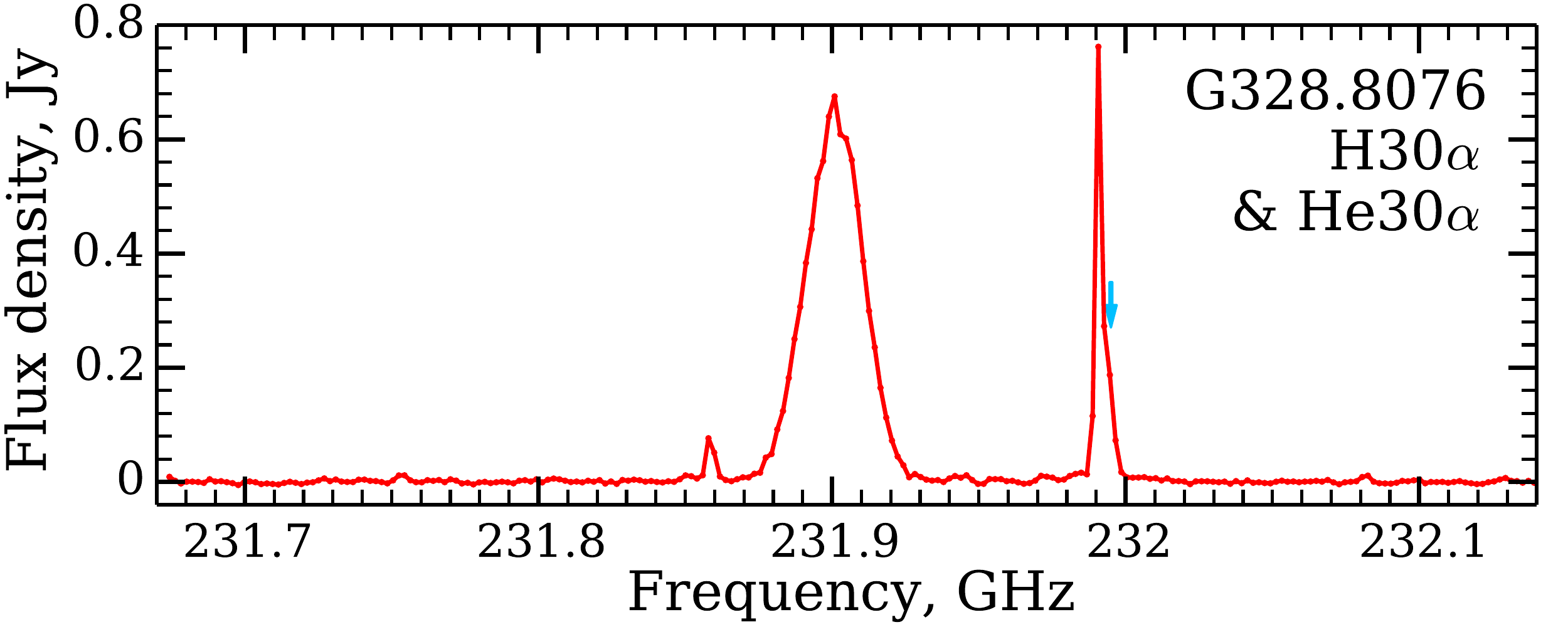} & \includegraphics[width=0.48\columnwidth]{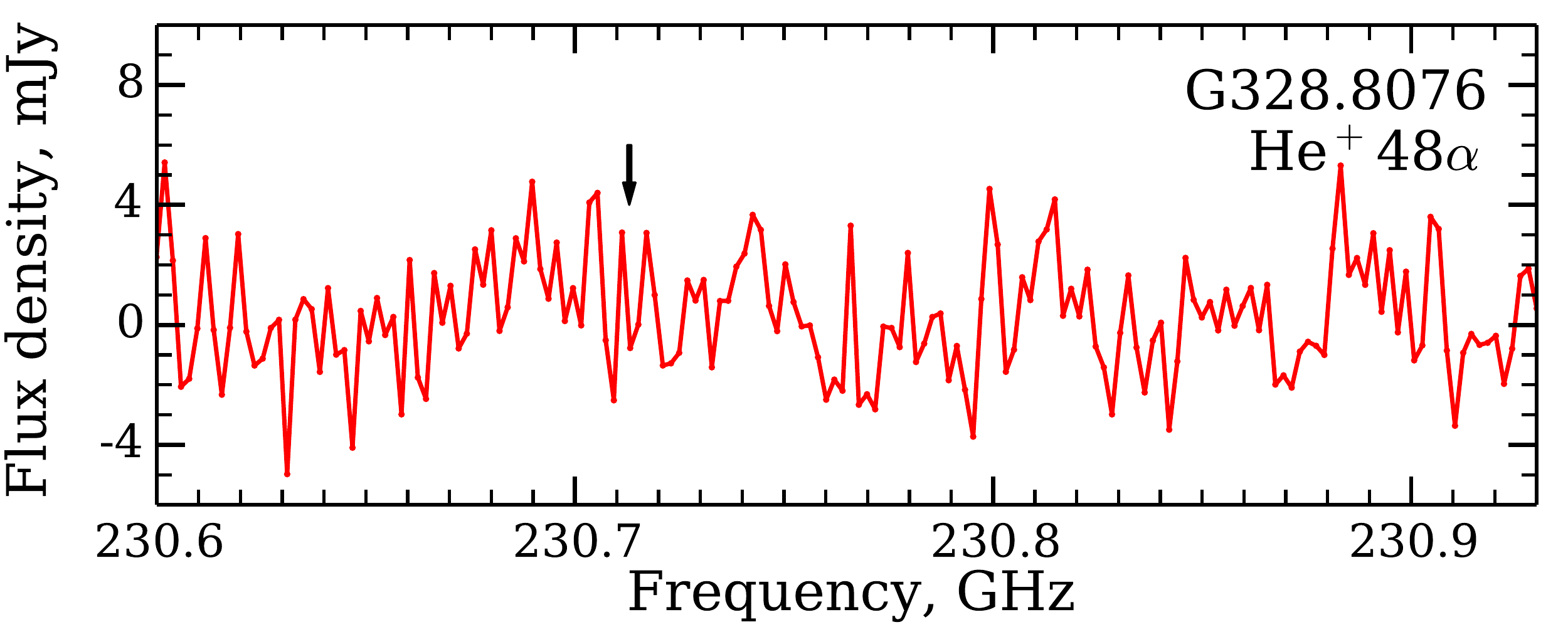}
\end{tabular}
\caption {Spectra of the brightest sub-regions of the HII regions G330.9536, G332.8254, Sgr B2(M) and G328.8076. The spectra are given in the rest frame of their respective sources. Left panels: The strong wide peaks at 231.9 GHz are H30$\a$ lines. The weak wide peaks at 231.995 GHz marked with blue arrow are He30$\a$ lines. The rest of the features in the spectra are molecular lines, which are much narrower than H30$\alpha$ and He30$\alpha$. The strong and narrow feature just on the left from G328.8076's He30$\a$ line is a molecular line. Right panels: Black arrows mark the central frequency of the undetected He$^+$48$\a$ line. The level of spectral noise is $\sim 0.5$ mJy per $10 \, \kms$ channel for G330.9536 and G332.8254, and $2$ mJy per $10 \, \kms$ channel for Sgr B2(M) and G328.8076, so most of the features for example in G332.8254 are molecular lines, not noise.  All spectra here are given for extraction 1 of their respective sources (see Figure \ref{fig:regions}).}
\label{fig:cycle4}
\end{figure*}

\begin{figure*}
\vspace{-0.0cm}
\centering
\begin{tabular}{cccc}
\includegraphics[width=0.23\columnwidth]{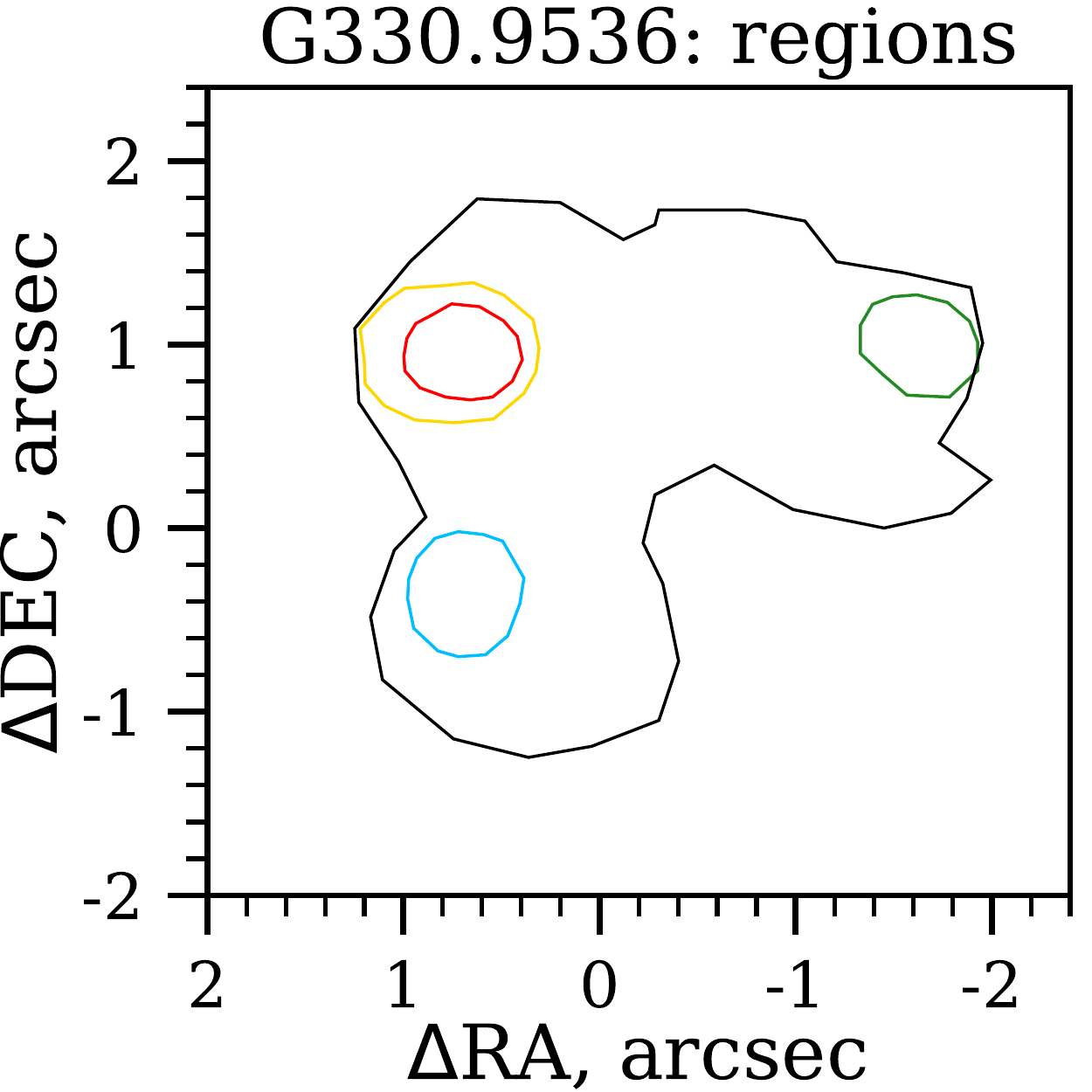}   & 
\includegraphics[width=0.23\columnwidth]{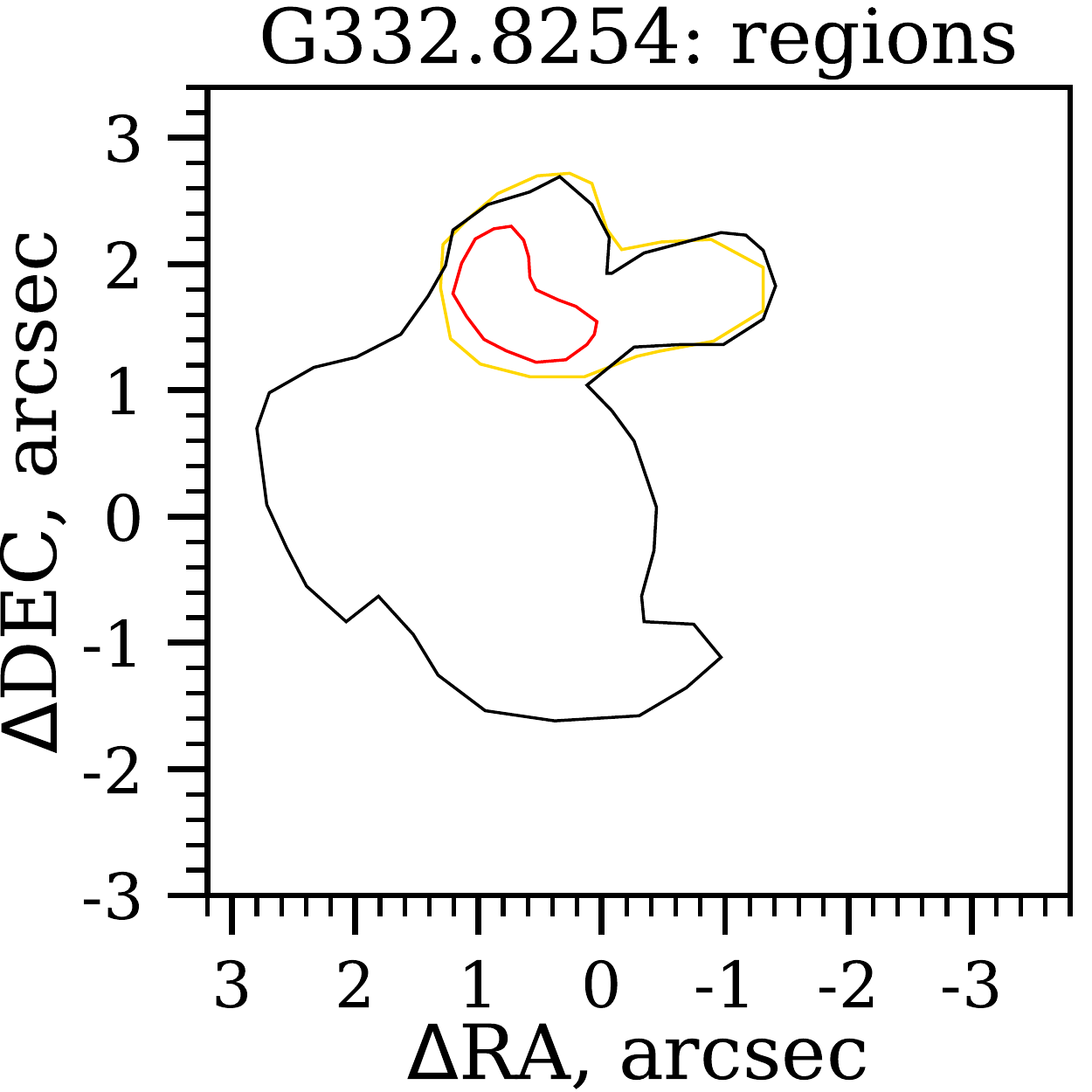}   &
\includegraphics[width=0.23\columnwidth]{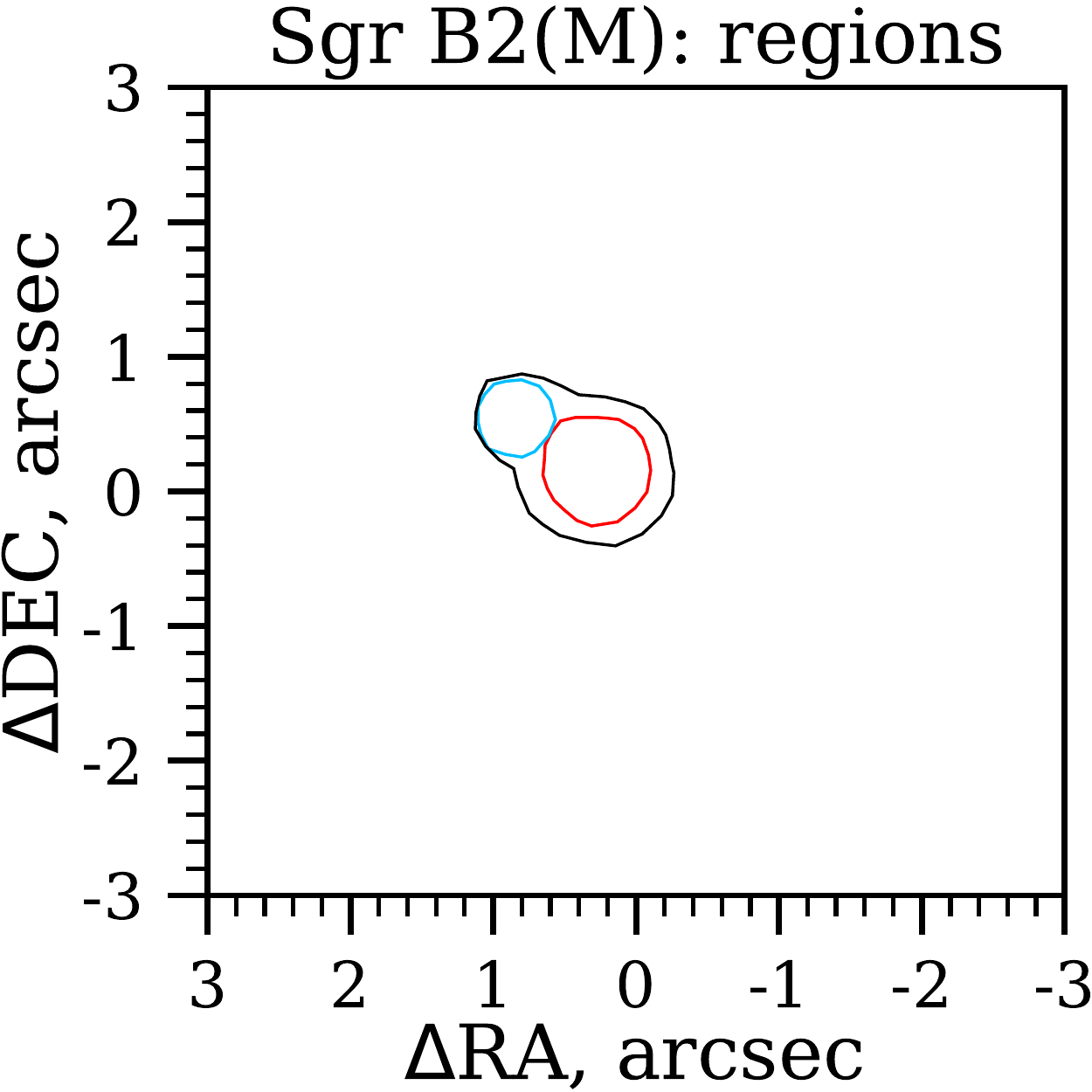} & 
\includegraphics[width=0.23\columnwidth]{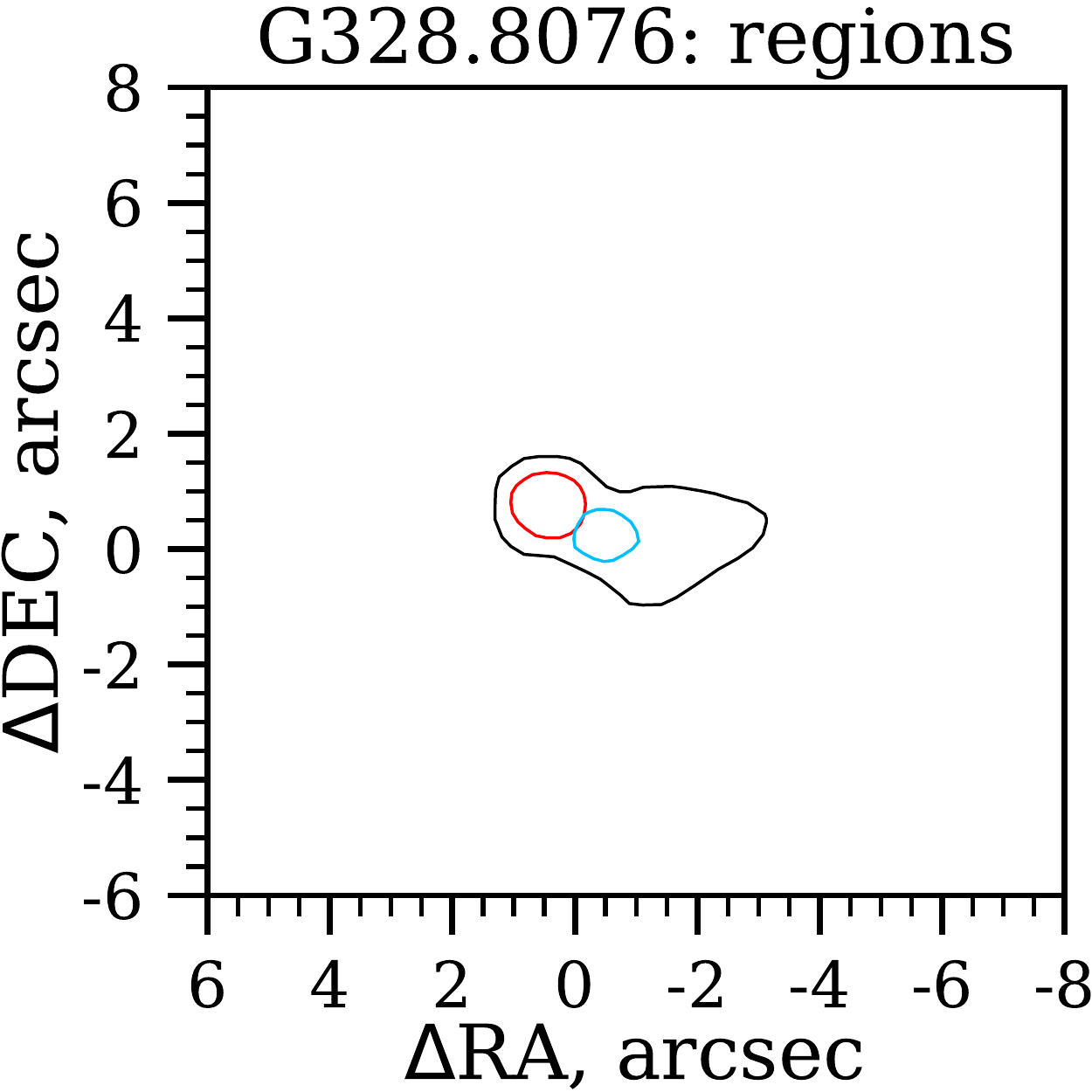}
\end{tabular}
\caption {Sub-regions of the four observed HII regions for which we extract spectra and calculate the intrinsic EUV spectral parameters. Sub-region 1 in all targets is marked in red. Yellow marks sub-region 1a which encloses sub-region 1. Sub-region 2 and 3 are marked in blue and green respectively. The sub-region ``tot$_\mathrm{H30\a}$'' enclosing most of the H30$\a$ emission is marked in black.}
\label{fig:regions}
\end{figure*}

We selected several of the H30$\a$-brightest sub-regions, as well as identified large regions called ``tot$_\mathrm{H30\a}$'' encapsulating most of the H30$\a$ emission. Extraction regions are shown in Figure \ref{fig:regions}. While the extraction region ``tot$_\mathrm{H30\a}$'' encapsulates most of the H30$\a$ emission, it does not necessarily contain most of the continuum emission. 
Some of the targets contain multiple non-overlapping sub-regions, each of which can be considered as an independent probe of the underlying EUV spectrum. We extract spectra within these sub-regions and analyze them. An example spectrum for each target is shown in Figure \ref{fig:cycle4}.

Computation of the velocity-integrated line flux can be achieved by integrating under the recombination line. However, in order to obtain accurate values we need to remove contributions of molecular lines. Molecular lines are easily identifiable in most cases by their much narrower width.  
To remove contributions of molecular lines to velocity-integrated line flux of the recombination lines we fitted the spectra with a combination of Gaussian and Voigt profiles and then used the fit to extract central velocity ($V$) of the H30$\a$ and He30$\a$ 
lines, their width ($\D V$) and the velocity-integrated line flux ($[S\Delta V]_\mathcal{L}$).  In the cases when the HeI line is positioned on the wing of the HI line, we fit the HI line first, then subtract its contribution from the HeI line and fit the HeI line with the Gaussian or a combination of Gaussian and Voigt profiles in the same way as the HI line.

The HeII recombination line was not detected in any of our targets. This is expected as the number of $h\nu > 54.4\eV$ photons is predicted to be quite low in SF regions (see below). To set the limit on $[S\Delta V]_\mathrm{He^+48\a}$ we assume that $V_\mathrm{He^+48\a} = V_\mathrm{H30\a} \simeq V_\mathrm{He30\a}$ and $\D V_\mathrm{He^+48\a} = \D V_\mathrm{He30\a}.$ These imply that HII, HeII and HeIII Stromgren spheres are at rest with respect to each other and that the bulk motions of the gas, which dominate the line recombination width (see below) within HeII and HeIII Stromgren spheres are similar.
The derivation of the He$^+48\a$ flux limit is also complicated by the presence of strong molecular features on both sides of the line at about $\pm 50 \, \kms.$ Thus instead of integrating the expected velocity range around the central frequency we use the width and the velocity of He30$\a$ lines to obtain the expected He$^+$48$\a$ profile which we then fit under the uncontaminated spectral data points around the central frequency. 

The results of the measurements and calculations are given in Table \ref{summary}.
The line-of-sight thermal velocities of H and He atoms at the characteristic temperature of HII regions of $T \sim 10^4 \, \K$ are $\bar{V}_z^\mathrm{H} \sim 9 \, \kms$ and $\bar{V}_z^\mathrm{He} \sim 4.5 \, \kms,$ a factor of a few less that the observed line widths. This is due to the fact that a considerable part of the motion reflected in the recombination lines' widths is due to the bulk motion of the gas (inflow, outflow, rotation), blending of components, etc. 

We present continuum flux densities within the extraction regions ($S_\mathrm{cont}$) in Table \ref{summary}. The continuum flux densities can be used in conjunction with the recombination line fluxes to estimate the temperature within the HII regions. To do this one assumes that most continuum radiation is coming from the free-free emission. 
The temperature estimates for the observed extraction regions obtained this way are between 5,000 K to 12,000 K, in agreement with the temperatures expected within HII regions. A caveat in these estimates, however, is that the continuum emission in such dusty sources as our HII regions is a combination of free-free emission and dust emission and they cannot be separated without extensive multi-wavelength coverage. Thus the temperatures estimated as discussed above are highly uncertain, and we opt to omit discussing them in detail.

We estimate the uncertainties of the line width and the central line velocity at $\d V = 2.5 \, \kms.$
The uncertainties on the velocity-integrated line flux are $\d [S\Delta V] \sim 0.01$ Jy $\kms$ in the Cycle 4 and $\d [S\Delta V] \sim 0.1$ Jy $\kms$ in the Cycle 2 observations.
The accuracy of determining $[S\Delta V],$ $V$ and $\Delta V$ is dominated by molecular lines contamination, which is hard to quantify.

For G330.9536 and G332.8254 we estimate the combined uncertainty on $[S\Delta V]_\mathrm{H30\a}$ at 3\%, the uncertainty on $[S\Delta V]_\mathrm{He30\a}$ at 10\% and the uncertainty on $[S\Delta V]_\mathrm{He^{+}30\a}$ at 30\%. For G328.8076 and Sgr~B2(M) we estimate the combined uncertainty on $[S\Delta V]_\mathrm{H30\a}$ at 5\%, the uncertainty on the limit on $[S\Delta V]_\mathrm{He30\a}$ at 30\% and the uncertainty on $[S\Delta V]_{He^{+}48\a}$ at 50\%. We calculate each velocity-integrated line flux using the procedure described earlier in this Section. In the spectrum of G328.8076 we remove all points belonging to the strong and narrow molecular feature just to the left from the He30$\a$ line and fit the Gaussian-like profile to the remaining points. The presence of strong molecular feature on top  of the He30$\a$ line in G328.8076 and the fact that the estimated $[S\Delta V]_\mathrm{He30\a}$ for all sub-regions are comparable with the observational uncertainties make the derived values of $\gamma$ for this target unreliable.

The uncertainties in $[S\Delta V]_\mathcal{L}$ directly translate into uncertainties on $\EM_x$ via eq. \ref{SdVtoEM}, into $Q$'s via eq. \ref{eq:EMtoQ} and to $L_\nu,$ $\gamma$ and $\gamma_2$ via eq. \ref{eq:lnu}, implying uncertainty of $\sim 4$ \% on $\gamma$ for the Cycle 4 targets and of $\sim 10\%$ on Sgr B2(M). Since the equations are approximate, we expect the true uncertainties on the derived spectral parameters to be $\sim 20 \%.$
We are unable to provide a meaningful limit on $\gamma_2$ for G328.8076 and Sgr B2(M).

\bigskip

\section{Discussion}
\label{sec:results}

\begin{figure*}
\vspace{-0.0cm}
\centering
\begin{tabular}{cc}
\includegraphics[width=0.48\columnwidth]{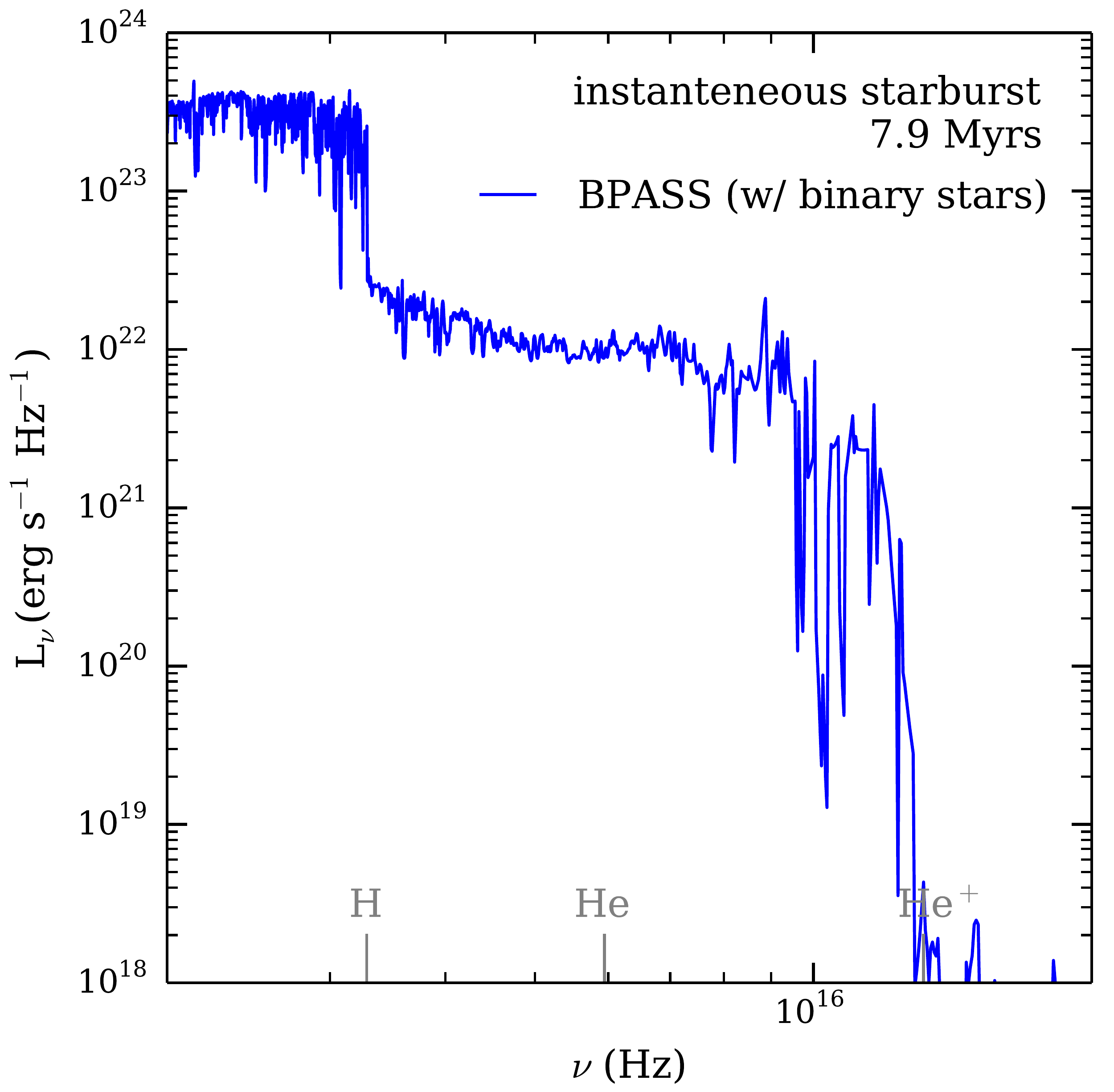}   & 
\includegraphics[width=0.48\columnwidth]{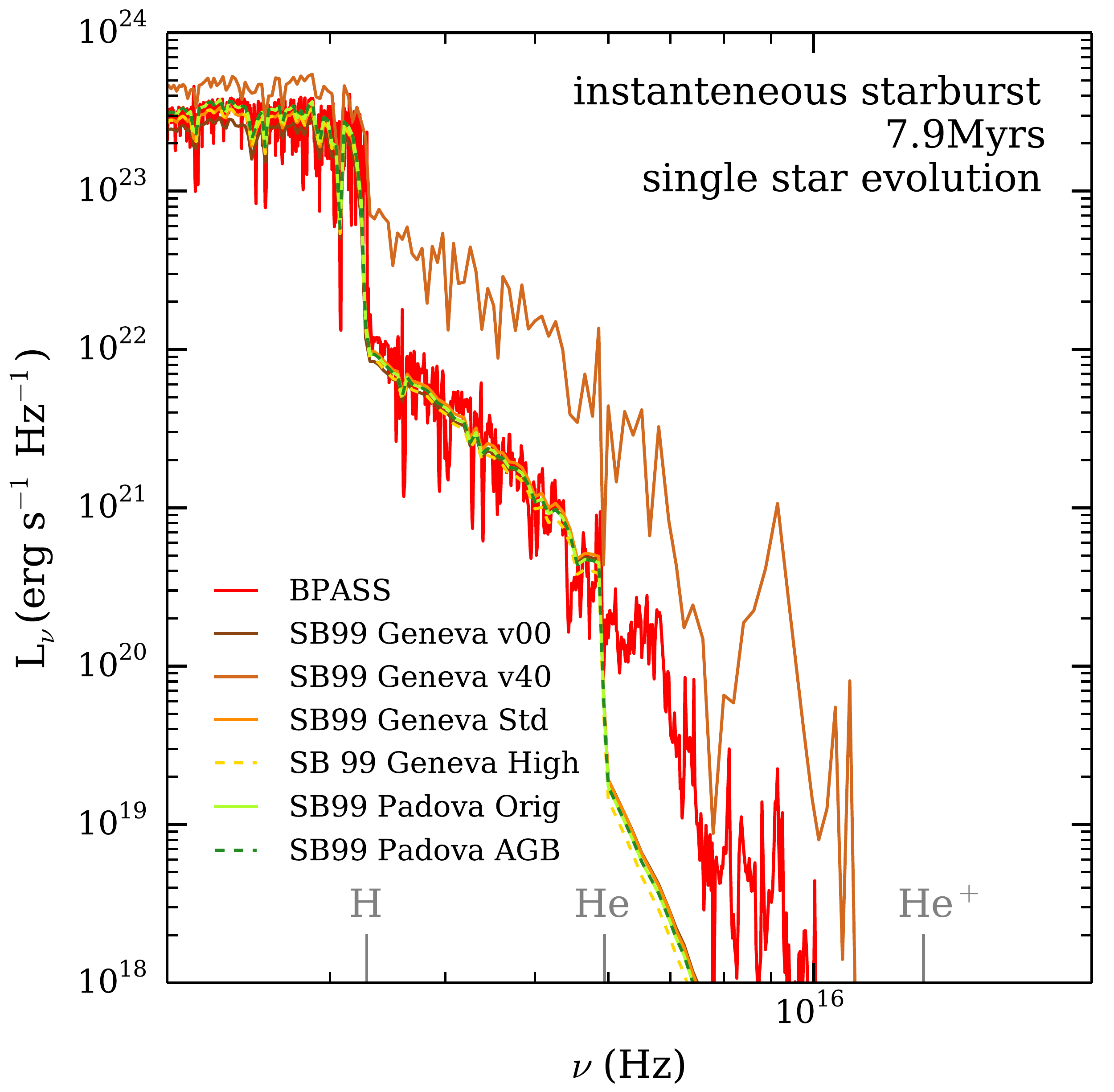}
\end{tabular}
\caption{Example spectra of an instantaneous starburst of $10^3 \Msun$ at 7.9 Myrs in the models with only single star evolution --  BPASS v2.2.1 single star and various models used in Starburst99 v7.0.1 (right panel) and with binary populations -- BPASS v2.2.1 (left panel). In all cases the metallicity is set to $Z=0.02,$ except for Starburst 99 models Geneva v00 and v40 for which only $Z=0.014$ was available.}
\label{fig:models} 
\end{figure*}

In Table \ref{summary} we present derived values for the volume emission measures of HII, HeII and HeIII (eq. \ref{SdVtoEM}), the production rate of EUV photons $Q_0:$ $h\nu>h\nu_0=13.6$ eV, $Q_1:$ $h\nu>h\nu_1=24.6$ eV, $Q_2:$ $h\nu>h\nu_2=54.4$ eV (eq. \ref{eq:EMtoQ}) and the inferred  parameters of the underlying EUV spectra $L_\nu,$ $\gamma$ and $\gamma_2$ (eq. \ref{eq:lnu}) for each extraction region of G330.9536, G332.8254, Sgr B2(M) and G328.8076 (Figures \ref{fig:sources} and \ref{fig:regions}). We see that for all extraction regions the specific luminosity of EUV for $13.6 \, \eV \leq h \nu < 54.4 \, \eV$ scales close to $L_\nu \sim \nu^{-4.5 \pm 0.4},$ and steepens even more for higher energies $h \nu \geq 54.4$ eV to $\sim \nu^{-\gamma_2},$ where $\gamma_2>15$. The value of the high energy spectral slope is dominated by G330.9536 which has lower molecular line content and thus allows for much more stringent limit on $[S\D V]_\mathrm{He^+48\a}$ and thus on $\gamma_2.$

Above we quote averaged values of the exponents over the non-overlapping extraction regions of the observed HII regions excluding G328.8076.
The spectral slope $\gamma$ obtained for G328.8076 is $\sim 7.5$. However, the presence of strong molecular line features near He30$\a$ in its spectrum leads to large uncertainties and makes this estimate the least reliable. So we exclude this HII region from the calculation of $\langle \gamma \rangle$ and the following discussion. The 1 sigma uncertainties of the averaged spectral slope $\langle \gamma \rangle$ quoted above are dominated by the uncertainties of the model assumptions and not the standard deviation due to the averaging.
% is $0.3.$}

\begin{figure*}
\vspace{-0.0cm}
\centering
\begin{tabular}{cc}
\includegraphics[width=0.48\columnwidth]{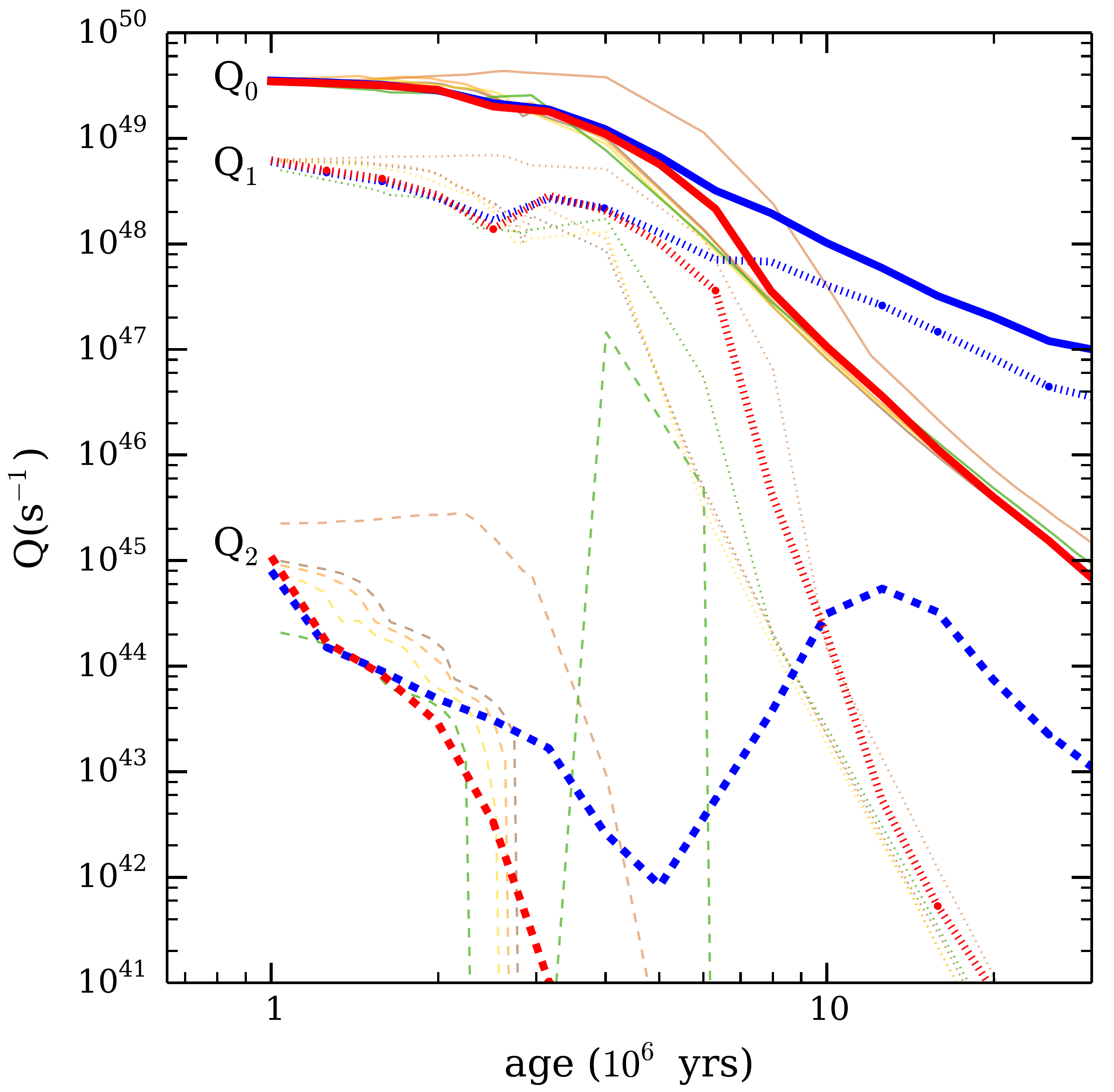}   & 
\includegraphics[width=0.48\columnwidth]{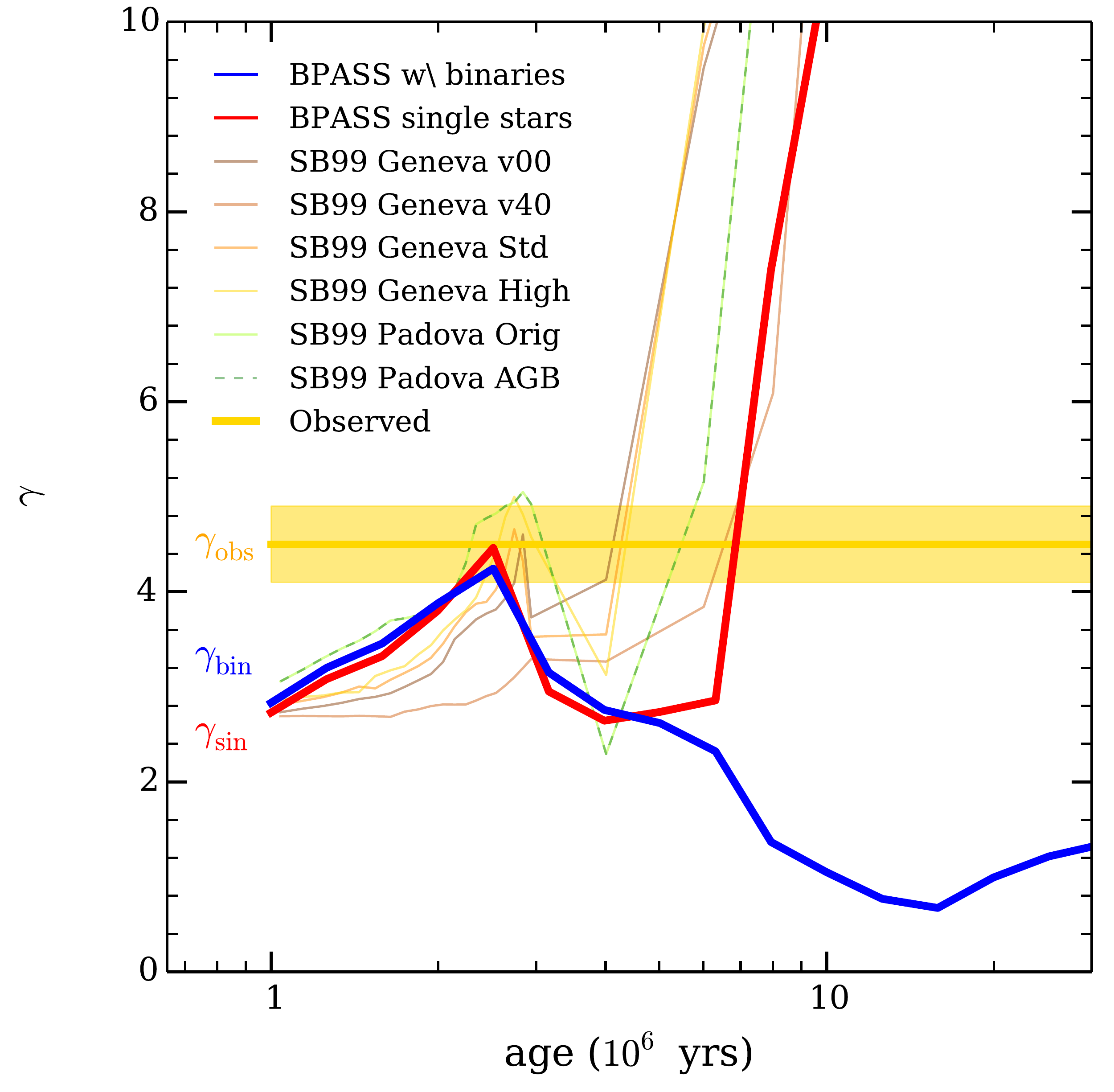}
\end{tabular}
\caption {Left panel: The evolution of the production rates of EUV photons with energies above the ionization thresholds of HI, HeI and HeII, i.e. $h\nu > 13.6$ eV ($Q_0$), $h\nu > 24.6$ eV ($Q_1$) and $h\nu > 54.4$ eV ($Q_2$), respectively, for instantaneous starburst scaled to $10^3 \Msun$ in the models with only single star evolution -- BPASS v2.2.1 single stars (red) and various Starburst99 v7.0.1 models and with binary populations -- BPASS v2.2.1 (blue). Right panel: Evolution of the slope ($\gamma$) of the EUV spectra of $L_\nu \sim \nu^{-\gamma},$ where $13.6 \eV \leq h \nu < 54.4 \eV$ for the same models. The values are obtained with eq. \ref{eq:Caa2}. Yellow line marks the average observed value of $\gamma = 4.5 \pm 0.4.$}
\label{fig:Qs.gammas}
\end{figure*}

We compare this observational result with model spectra of an instantaneous $10^3 \Msun$ starburst obtained with (i) Starburst99 v7.0.1 \citep{1999ApJS..123....3L,2014ApJS..212...14L} which assumes that all stars evolve as single stars only and (ii) the Binary Population and Spectral Synthesis code (BPASS) v2.2.1 \citep{2017PASA...34...58E,Stanway18} which can take into account binary population and the interactions between the stars such as common envelope evolution and mass transfer. The BPASS can be set to evolve all stars as single stars, i.e. with binary interaction switched off.
For illustration we present spectra produced by the two codes at 7.9 Myrs (Figure \ref{fig:models}). Left panel shows the spectra produced by BPASS with binary interaction included. The right panel show the spectra for single star evolution models produced by BPASS with binary interaction switched off and Starburst99 with the six evolutionary tracks -- Geneva track with zero rotation (v00), Geneva tracks with $V=0.4$ rotation (v40), Geneva tracks with standards mass loss, Geneva tracks with high mass loss, original Padova tracks, Padova tracks with AGB stars. For description and the appropriate references see \citealt{2014ApJS..212...14L}. In all cases the metallicity is set to $Z=0.02,$ except for Starburst 99's Geneva v00 and v40 tracks for which only $Z=0.014$ is available. We scale the simulations to $10^3 \Msun$ so that the production rate of $h\nu > 13.6 \, \eV$ photons ($Q_0$) predicted by the models is similar to the observed values for our extraction regions (see left panel in Figure \ref{fig:Qs.gammas} and Table \ref{summary}). The time range on the figure is set to span the lifetime of a single OB star, which is $< 30$ Myrs.

Figure \ref{fig:Qs.gammas} (right panel) shows the evolution of the slope ($\gamma$) of the EUV spectra $L_\nu \sim \nu^{-\gamma_*},$ where $13.6 \eV \leq h \nu < 54.4 \eV,$ for Starburst99 v7.0.1 (SB99) and BPASS v2.2.1 models.
We find that the observed spectral slope $\g \simeq 4.5 \pm 0.4$ differs from model predictions. There are no precise enough methods of estimating the absolute age of each SF regions, thus we compare the differences between the measured EUV spectral slope and the model for all ages younger than a typical lifetime of an OB star.
For an instantaneous starburst of $<7 \times 10^6$ Myrs in age the observationally inferred slopes $\langle \gamma \rangle \simeq 4.5$ are consistently steeper than in either of the models $\gamma_\mathrm{sin} \simeq \gamma_\mathrm{bin} \simeq 3$ by a factor of $\sim$1.5, meaning that the observed spectrum is considerably softer than the theoretical one. The values 3 and 4.5 may seem close enough and one may be tempted to declare an agreement between the observations and the simulations. However it is crucial to stress that we are comparing the values of exponents and that the spectra scaling as $\nu^{-3}$ or $\nu^{-4.5}$ are substantially (4 sigma) different from each other. Thus we conclude that the observations do not agree with the models. For an instantaneous starburst of $>7 \times 10^6$ Myrs in age the observed spectral slope is in between the two model ones -- steeper than $\gamma_\mathrm{bin}$ and less steep than $\gamma_\mathrm{sin}$.

\begin{longrotatetable}
\begin{deluxetable*}{cccccccccccccccccc}
\tabletypesize{}
\tablecaption{\bf{Parameters of the observed star forming regions} \label{summary}  }
\tablewidth{700pt}
\tabletypesize{\scriptsize}
\tablehead{
\colhead{Reg.} & \colhead{$S_{\rm cont}$} 
& \colhead{$V_{\rm HI}$} & \colhead{$\Delta V_{\rm HI}$} & \colhead{$[S\D V]_{\rm H30\a}$} & \colhead{$\frac{\EM_{\rm HII}}{10^{61}}$} & \colhead{$\frac{Q_0}{10^{48}}$} 
& \colhead{$V_{\rm HeI}$ } & \colhead{$\Delta V_{\rm HeI}$ } & \colhead{$[S\D V]_{\rm He30\a}$ } & \colhead{$\frac{\EM_{\rm HeII}}{10^{60}}$} & \colhead{$\frac{Q_1}{10^{47}}$}
& \colhead{$[S\D V]_{\rm He^{+}48\a}$ } & \colhead{$\frac{\EM_{\rm HeIII}}{10^{58}}$} & \colhead{$\frac{Q_2}{10^{46}}$} 
& \colhead{$\frac{L_0}{10^{23}}$} & \colhead{$\gamma$} & \colhead{$\gamma_2$} 
\\
\colhead{name} & \colhead{Jy} & \colhead{$\kms$} & \colhead{$\kms$} & \colhead{Jy $\kms$} & \colhead{$\cm^{-3}$} & \colhead{$\sec^{-1}$}
& \colhead{$\kms$} & \colhead{$\kms$} & \colhead{Jy $\kms$} & \colhead{$\cm^{-3}$} & \colhead{$\sec^{-1}$}
& \colhead{Jy $\kms$} & \colhead{$\cm^{-3}$} & \colhead{$\sec^{-1}$}
& \colhead{$\frac{\erg}{\sec \, \Hz}$}& &
}
\startdata
\\
\multicolumn{18}{c}{\bf G330.9536} \\
1     & 1.43 & -90.0 & 42.5 & 146.70 & 3.91 & 10.12 & -92.3 & 41.5 & 8.75 & 2.33 & 6.34 & $<0.04$ & $<0.34$ & $<0.62$ & 1.86 & 4.6 & $>12$\\
1a    & 1.81 & -90.8 & 42.8 & 170.72 & 4.55 & 11.78 & -91.9 & 45.4 & 11.59& 3.09 & 8.40 & $<0.10$ & $<0.73$ & $<1.36$ & 2.34 & 4.4 & $>9$\\
2     & 0.77 & -92.8 & 30.6 &  62.72 & 1.67 & 4.33  & -93.3 & 32.9 &  5.50& 1.47 & 4.00 & $<0.03$ & $<0.19$ & $<0.34$ & 1.00 & 3.9 & $>21$\\
3     & 0.54 & -83.1 & 42.9 &  41.57 & 1.11 & 2.87  & -83.9 & 37.8 & 2.31 & 0.62 & 1.69 & $<0.07$ & $<0.05$ & $<0.10$ & 0.51  & 4.7 & $>20$\\
tot$_\mathrm{H30\a}$ & 7.01 & -89.7 & 35.2 & 485.04 & 12.93& 33.49 & -88.2 & 39.9 &34.36 & 9.17 & 24.94& $<0.13$ & $<0.97$ & $<1.79$ & 6.83 & 4.3 & $>20$\\
\\
\multicolumn{18}{c}{\bf G332.8254} \\
1     & 1.02 & -64.7 & 30.1 & 85.46  & 1.46 & 3.78 & -66.7 & 26.7 & 6.48 & 1.11  & 3.02 & $<0.08$ & $<0.39$ & $<0.72$ & 0.80  & 4.2 & $>6.5$\\
1a    & 1.98 & -63.7 & 33.4 &147.63  & 2.52 & 6.53 & -66.6 & 28.5 &10.25 & 1.75  & 4.76 & $<0.21$ & $<1.01$ & $<1.87$ & 1.31 & 4.4 & $>3.4$\\
tot$_\mathrm{H30\a}$ & 4.75 & -64.0 & 31.0 &275.78  & 4.71 & 12.20& -68.4 & 21.8 & 21.26& 3.63  & 9.87 & $<1.58$ & $<2.82$ & $<5.21$ & 2.61 & 4.2 & $>3.0$\\
\\
\multicolumn{18}{c}{\bf Sgr B2(M)} \\
1     &3.92  & 51.8  & 38.5 & 217.48 & 13.21& 34.21& 52    & 39   & 12.29 & 7.47 & 20.32& $<1.434$& $<24.79$& $<45$   & 6.07 & 4.7 & - \\
2     &0.43  & 55.2  & 40.9 & 32.42  & 1.97 & 5.10 & 55    & 41   & 1.83  & 1.11 & 3.02 & $<0.435$& $<7.52 $& $<14$   & 0.90  & 4.7 & - \\
tot$_\mathrm{H30\a}$ &5.49  & 51.7  & 39.8 & 268.10 & 16.28& 42.17& 52    & 40   & 12.71 & 7.72 & 21.00& $<2.33$ &$<40.26$ & $<75$   & 6.67 & 5.0 & - \\
\\
\multicolumn{18}{c}{\bf G328.8076} \\
1     & 0.25 & -51.4 & 29.2 & 20.12  & 2.43 & 6.29 & -50   & 30   & 0.35 & 0.42  &  1.15 & $<0.07$ & $<2.41$ & $<4.4$ & 0.49 & 6.7 & - \\
2     & 0.21 & -57.2 & 36.2 & 18.33  & 2.21 & 5.72 & -57   & 36   & 0.15 & 0.19  &  0.51 & $<0.08$ & $<2.65$ & $<4.9$ & 0.25 & 8.0 & - \\
tot$_\mathrm{H30\a}$ & 0.94 & -54.1 & 32.3 & 72.95  & 8.80 & 22.79& -54   & 32   & 0.69 & 0.83  &  2.26 & $<0.34$ & $<11.81$& $<21.8$&1.11 & 7.8 & - \\
\\
\enddata
\tablecomments{Full set of observed parameters of the H30$\a,$ He30$\a$ and He$^+$48$\a$ recombination lines. Their central velocities ($V$), velocity-widths ($\D V$), velocity-integrated line fluxes ($[S\D V])$,
derived volume emission measures ($\EM$) (eq. \ref{SdVtoEM}), production rates of ionizing photons ($Q$) (eq. \ref{eq:EMtoQ}) and model parameters of the underlying EUV spectra ($L_0,$ $\gamma,$ $\gamma_2$) (eq. \ref{eq:lnu})
for compact HII regions G330.9536, G332.8254 and G328.8076 and starburst Sgr B2(M). Extraction regions are shown in Figure \ref{fig:regions}. Reg. name is the name of the extraction region, i.e. sub-region of the studies HII region. $S_\mathrm{cont}$ is continuum flux measured within the extraction region. For discussion of the uncertainties see Section \ref{sec:data2}.}
\end{deluxetable*}
\end{longrotatetable}

Here and further when referring single star evolution models we are primarily quoting values prodiced by BPASS single star. The Starburst 99 and BPASS single star evolution models predict similar results for $Q_0,$ $Q_1$ and $\gamma.$ The predictions for $Q_2$ vary substantially, however we cannot set any meaningful constraint on these values as they are below our sensitivity limit. Thus we do not discuss variations in $Q_2$ here.

Our observationally inferred EUV spectral slopes for HII regions coincide with the model ones for a $\sim 2.5$ Myr-old starburst. One may wonder whether this implies that all our HII regions are about the same age which is $\sim$ 2.5 Myrs. We rule out this possibility. The survey by \cite{2010MNRAS.405.1560M} identified G330.9536 as Ultra-Compact or Hyper-Compact HII region, thus its OB stars/clusters of stars are likely at the earliest stage of the evolution. G332.8254 and, the excluded from the discussion, G328.8076 are classified as Ultra-Compact HII implying that they are larger and thus likely older than G330.9536. Sgr B2(M) consists of several classical well-developed HII regions, placing it at a later evolutionary stage than any of the G-sources. Our HII regions span several stages of HII region evolution and of their embedded OB star population evolution and cannot be of the same age. This makes our finding that all of the EUV spectra have similar shapes to $L_\nu \sim \nu^{-4.5}$ even more puzzling. 

Let us now assume that the ages of all our SF regions are $> 7$ Myrs.
For simplicity we also assume that the primary difference between the single star evolution and binary evolution codes is the interacting binary fraction, which is zero for Starburst99 and BPASS single star and $\sim$60\% for massive $\sim 5 \Msun$ stars, exceeding $\sim 90$ \% for $M>16 \Msun$ stars in BPASS v2.2.1 (see Figure 1 of \citealt{Stanway18}). The latter is comparable with the 70\% estimated by \citep{2012Sci...337..444S} for O stars.
For such ages the observationally determined EUV spectral slope is always in between the two models, i.e. steeper than for the BPASS's binary evolution model and less steep than for the Starburst99 and BPASS single star models. This can imply that the binary fraction also has to be in between the binary fractions assumed in BPASS binary evolution and in Starburst99 or BPASS single star, i.e. lower than in BPASS binary evolution and greater than in Starburst99 or BPASS single star. Let us take as an example a 7.9 Myr-old SF region whose model spectra are presented in Figure \ref{fig:models}. At this age the model EUV spectral slope we can infer either from Figure \ref{fig:models} or from Figure \ref{fig:Qs.gammas} (right panel) is $\gamma_\mathrm{bin} \simeq 1.5$ and $\gamma_\mathrm{sin} \simeq 7.5$ compared to the observed $\gamma = 4.5,$ imply that we need to assume lower binary fraction than assumed in BPASS v2.2.1 (see Figure 1 of \citealt{Stanway18}).

EUV spectral slopes in extraction regions of Sgr B2(M) are consistently steeper than for the G-sources, which may be an indication of spectra steepening with age similar to the trend predicted by single star evolution models with its zero binary fraction, while BPASS with binary evolution predicts the spectral slope marginally decreasing at the later stages of evolution while mostly staying at $\gamma_\mathrm{BPASS} \sim 1.$  This again may serve as an indicator that the binary fraction in the SF regions is likely smaller than $60$ \% for massive $\sim 5 \Msun$ stars and exceeding $\sim 90$ \% for $M>16 \Msun$ stars in BPASS v2.2.1 (Figure 1 of \citealt{Stanway18}). Another possibility is that the steepening may be due to partial escape of EUV photons from the region. Our calculations are based on the assumption that all EUV photons are captured and reprocessed into lower energy recombination line photons by H and He. This is a particularly good approximation for ultra-compact and hyper-compact HII regions, but may lead to less accurate estimated for evolved HII regions such as Sgr B2(M) from which photons can escape more easily.

In the analysis above we treat non-overlapping extraction regions in each of the sources as independent probes of spectra of SF regions.
We have three non-overlapping sub-region in G330.9536, one in G332.8254 and two in Sgr B2(M). The presence of several SF clusters is typical for an HII region. Although different clusters are a part of the same HII region, and thus have similar chemical composition, environmental properties, and they are of similar, but not necessary the same, age, they nonetheless can be considered fairly independent probes of EUV spectra of star formation. It is believed that sub-regions within HII regions are dominated by either one OB star or by a cluster of a few OB stars. In this case, we do not expect that binary fractions within each non-overlapping sub-region are correlated. Thus sub-regions within HII regions can be treated as independent.

\begin{figure*}
\vspace{-0.0cm}
\centering
\begin{tabular}{cc}
\includegraphics[width=0.48\columnwidth]{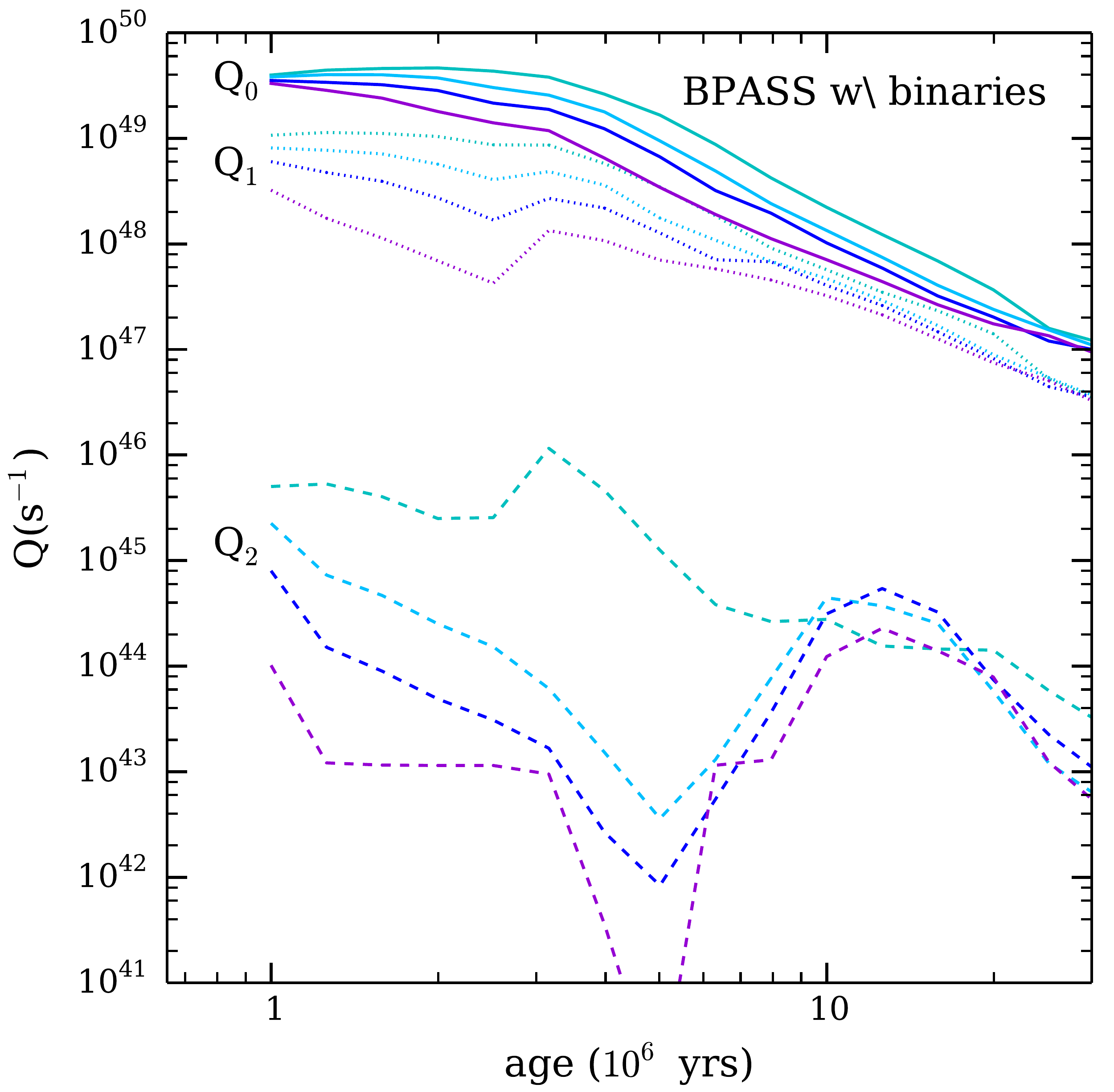}   & 
\includegraphics[width=0.48\columnwidth]{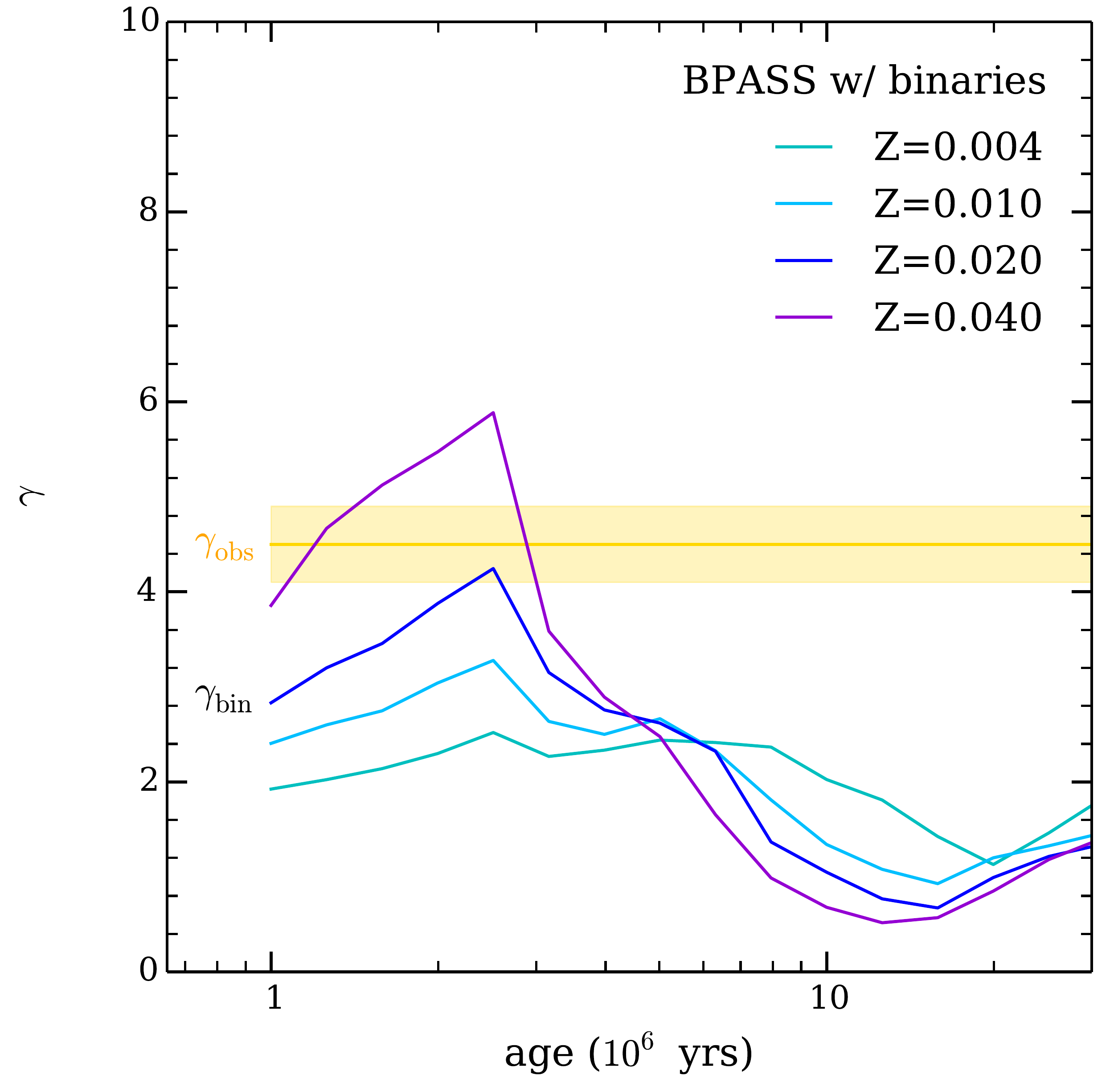}
\end{tabular}
\caption{Left panel: The evolution of the production rates of EUV photons with energies above the ionization thresholds of HI, HeI and HeII, i.e. $h\nu > 13.6$ eV ($Q_0$), $h\nu > 24.6$ eV ($Q_1$) and $h\nu > 54.4$ eV ($Q_2$), respectively, as a function of metallicity for instantaneous starburst scaled to $10^3 \Msun$ in the model with binary populations produced by BPASS v2.2.1. Right panel: Evolution of the slope ($\gamma$) of the EUV spectra of $L_\nu \sim \nu^{-\gamma},$ where $13.6 \eV \leq h \nu < 54.4 \eV$ as a function on metallicity for the same models model with binary populations produced by BPASS v2.2.1. The values are obtained with eq. \ref{eq:Caa2}. Yellow line marks the average observed value of $\gamma = 4.5 \pm 0.4.$}
\label{fig:metal} 
\end{figure*}

In Figure \ref{fig:metal} we present the evolution of the production rates of EUV photons with energies above the ionization thresholds of HI, HeI and HeII, i.e. $h\nu > 13.6$ eV ($Q_0$), $h\nu > 24.6$ eV ($Q_1$) and $h\nu > 54.4$ eV ($Q_2$), respectively, and the evolution of the slope ($\gamma$) of the EUV spectra of $L_\nu \sim \nu^{-\gamma},$ where $13.6 \eV \leq h \nu < 54.4 \eV$ as a function of metallicity. The variations of spectral parameters are obvious, and yet it is still clear that the metallicity alone cannot account for the discrepancy of the model and the observed spectral properties.

\section{Conclusion}

We observed a sample of four HII regions spanning various stages of HII region evolution with ALMA in the mm recombination lines of HI, HeI and HeII -- H30$\a$, He30$\a$ and He$^+$48$\a,$ respectively, -- to derive the shape of the underlying EUV spectra. We find that irrespective of the evolutionary stage of the HII region the spectra are surprisingly similar to each other and scale with frequency as $L_\nu \sim \nu^{-4.5},$ for $13.6 \eV \leq h\nu < 54.4 \eV$ (Table \ref{summary}). We compared this result with model spectra obtained with population synthesis codes Starburst99 v7.0.1 and BPASS v2.2.1 (Figure \ref{fig:Qs.gammas}). Starburst99 and BPASS v2.2.1 single star assumes zero binary fraction, and BPASS v2.2.1 with bonary interaction assumes that about 60\% of $5 \Msun$ stars are in binaries and with the fraction rising to above 90 \% for $16 \Msun$ and higher.
We find that the observed spectral slope $\g \simeq 4.5 \pm 0.4$ differs from both model predictions. Because we do not know the exact ages of our SF regions, we compare both models with the data for all ages $<30$ Myrs.
If we assume that the ages of the observed SF regions are greater than $\sim$7 Myrs, then the observationally determined EUV spectral slope lies between the slope obtained with the two population synthesis codes. This would imply that the binary fraction within our HII regions must be substantially lower than assumed in BPASS v2.2.1 (see Figure 1 in \citet{Stanway18}). 

The technique demonstrated here is a unique probe of EUV spectra, as it is free from uncertainties of extinction corrections and trace ion abundances. The mm/submm recombination lines of HI, HeI and HeII are reliable probes of the EUV spectra inside SF populations allowing to determine the production rate of EUV ionizing photons above ionization thresholds of HI, HeI and HeII and reconstruct the properties of the EUV spectra.
In this work we applied the technique to Galactic HII regions, but it can be extended to galaxies. Among its most natural and powerful applications is the study of ultra-luminous infrared galaxies (see SM13) and star formation in the distant universe, which refers the period after the end of Dark Ages ($z\sim 7 - 15$) and up until the Cosmic Noon ($z\sim 2$). The insight into star formation in the distant universe can be obtained from local analogues for the distant galaxy population \citep{2005ApJ...619L..35H,2014MNRAS.439.2474S}. The lower metallicity of the gas from which the stars are formed in these galaxies would result in formation of more massive stars and correspondingly modified EUV spectral properties. There is a long-standing question of exactly which objects -- early galaxies or early active nuclei -- re-ionize the Universe and whether the observed number counts of these objects are consistent with the re-ionization requirements (see e.g. \cite{Dayal2020}). If the EUV spectrum of low-metallicity SF regions is softer than what was assumed in the previous re-ionization calculations and models, then it would make it even more difficult for early stars to re-ionization the Universe.

The technique demonstrated here opens a new avenue in studying the EUV spectra as it allows us to derive reliable observational constraints that can be used to test the predictions of  various star formation models in a way that was previously not possible, providing critical insight into photon production rates at $\l \leq 912 \mathring{\mathrm{A}}.$ This can serve as calibration to starburst synthesis models and to shed light on properties of star formation in distant universe and the properties of ionizing radiation during reionization. 

\bigskip

\section*{Acknowledgements}
We are grateful to Nick Scoville for co-writing the ALMA proposals and collaboration on the initial stages of the paper, to J.J. Eldridge, Susan Clark, David Guszejnov and to the anonymous referee for their thoughtful comments which helped improving the manuscript.

LM stipend at the IAS is provided by the Friends of the Institute for Advanced Study. A part of this work was conducted while LM was supported by SOS NRAO program and as a Groce Fellow at Caltech. LM is grateful to Dr. David and Barbara Groce for their kindness and support.

Part of this research was carried out at the Jet Propulsion Laboratory, California Institute of Technology, under a contract with the National Aeronautics and Space Administration.

This paper makes use of the following ALMA data:
  ADS/JAO.ALMA\#2013.1.00111.S and ADS/JAO.ALMA\#2016.1.01015.S ALMA is a partnership of ESO (representing
  its member states), NSF (USA) and NINS (Japan), together with NRC
  (Canada) and NSC and ASIAA (Taiwan) and KASI (Republic of Korea), in
  cooperation with the Republic of Chile. The Joint ALMA Observatory is
  operated by ESO, AUI/NRAO and NAOJ.

The National Radio Astronomy Observatory is a facility of the National Science Foundation operated under cooperative agreement by Associated Universities, Inc.

\bibliography{HIIregs_bibliography}{}
\bibliographystyle{aasjournal}

\appendix
\section{Ionizing photon count, volume emission measures, and EUV spectra}\label{app:ion}
%%%%%%%%%%%%%%%%%%%%%%%%%%%%%%%%%%%%%

We use standard notation for the production rates of the ionizing photons above the ionization thresholds of HI, HeI and HeII, see eq. \ref{Qs}.
We assume that EUV spectrum in SF regions can be represented by broken power-law
\be\label{eq.app:lnu}
 L_\nu = \left\{
 \ba{ll}    
 \size L_0 (\nu/\nu_0)^{-\gamma},    & \quad \nu_0 \leq \nu < \nu_2\\
 \size L_2 (\nu/\nu_2)^{-\gamma_2},& \quad \nu \geq \nu_2.
 \ea\right.
\ee 
Here $L_0, \, \gamma, \, L_2$ and $\gamma_2$ are constants, and $L_2= L_0 (\nu_2/\nu_0)^{-\gamma}$ to satisfy continuity at $\nu=\nu_2.$ 
Substituting relations (\ref{eq.app:lnu}) into eq. \ref{Qs} we have
\be
    && Q_0 = \frac{L_0/h}{\gamma} \left[ 1- \left(\frac{\nu_2}{\nu_0}\right)^{-\gamma} \right] +Q_2, \qquad 
    \label{eq:QCa} Q_1 = \frac{L_0/h}{\gamma} \left[ \left(\frac{\nu_1}{\nu_0}\right)^{-\gamma}- \left(\frac{\nu_2}{\nu_0}\right)^{-\gamma} \right] +Q_2, \\
    \nonumber && Q_2 = \frac{L_0/h}{\gamma_2} \left(\frac{\nu_2}{\nu_0}\right)^{-\gamma}.
\ee
We can solve this system numerically for $C,$ $\gamma$ and $\gamma_2.$ The solution can be written analytically if we make used of the fact that $Q_2 \ll Q_0$ and $Q_2 \ll Q_1.$ We get
\be\label{eq:Caa2}
    \frac{Q_1}{Q_0} \simeq \frac{\nu_1^{-\gamma}-\nu_2^{-\gamma}}{\nu_0^{-\gamma}-\nu_2^{-\gamma}}, \qquad 
    L_0/h \simeq \frac{\gamma Q_0}{1-(\nu_2/\nu_0)^{-\gamma}}, \qquad
    \gamma_2 = \frac{L_0/h - \gamma Q_0 }{Q_2} = \gamma \frac{Q_0}{Q_2} \frac{1}{(\nu_2/\nu_0)^{\gamma}-1}.
\ee 

To derive the ionizing photon production rates Q from the observed EM's we assume (i) that the gas is in a radiation equilibrium, so that the number of recombinations of each ion species is equal to the number of ionizations producing this ion species per unit time and (ii) that all ionizing photons are used within the region and none escapes. Additionally, in order to obtain analytical solutions, (iii) we going to disregard the possibility that high energy photons can be stolen by species with lower ionizing threshold (for detailed treatment see e.g. \citealt{ost06}). Thus we have: 
\be  
    Q_0\simeq \a^\mathrm{HI}_B {\rm EM}_\mathrm{HII}, \qquad
    Q_1\simeq \a^\mathrm{HeI}_B {\rm EM}_\mathrm{HeII}, \qquad \label{eq:QEM}
    Q_2 \simeq \a^\mathrm{HeII}_B {\rm EM}_\mathrm{HeIII},
\ee
where
\be\label{eq:aBs}
  \size \a_B^\mathrm{HI}   = 2.59 \times 10^{-13} \, \cm^3 \, \sec^{-1}, \quad  &
  \size \a_B^\mathrm{HeI} = 2.72 \times 10^{-13} \, \cm^3 \, \sec^{-1}, \quad &
  \size \a_B^\mathrm{HeII}= 1.85 \times 10^{-12} \, \cm^3 \, \sec^{-1}
\ee
are recombination coefficients of HI, HeI and HeII to all $\mathrm{n}\geq 2$ at $T=10^4$K.

Combining eqs (\ref{eq:QEM}) and (\ref{eq:Caa2}) we get
\be
    && \label{a} \frac{2.21^{\gamma}-1}{4^{\gamma}-1} \simeq 0.105 \left(\frac{{\rm EM}_{\rm HeII}}{10^{60} \, \cm^{-3}}\right) \left(\frac{{\rm EM}_{\rm HII}} {10^{61} \, \cm^{-3}}\right)^{-1}\\
    && L_0 \simeq 1.716 \times 10^{22} \frac{\erg}{\sec \, \Hz} \left(\frac{{\rm EM}_{\rm HII}}{10^{61} \, \cm^{-3}}\right) \frac{\gamma}{1-0.25^{\gamma}}\\
    && \label{a_2} \gamma_2 \simeq 1.4 \gamma \times 10^{2} \left(\frac{{\rm EM}_{\rm HII}}{10^{61} \, \cm^{-3}}\right) \left(\frac{{\rm EM}_\mathrm{HeIII}}{10^{58} \, \cm^{-3}}\right)^{-1} \left[4^{\gamma}-1\right]^{-1}\\
    && L_2 = 1.716 \times 10^{22} \frac{\erg}{\sec \, \Hz}  \left(\frac{{\rm EM}_{\rm HII}}{10^{61} \, \cm^{-3}}\right) \frac{\gamma}{4^\gamma-1}.
\ee 
Using eq. \ref{a} we can solve for $\gamma$, which we then substitute it into the rest of the equations to find $L_0, \, L_2$ and $\gamma_2,$ 
thus obtaining the intrinsic EUV spectra inside the region.

\end{document}